\documentclass[ALICE,manyauthors]{cernphprep}
\usepackage[comma,square,numbers,sort&compress]{natbib}
\usepackage[colorlinks,linktocpage]{hyperref}
\usepackage{lineno}
\usepackage{xspace}
\usepackage{float}
\usepackage[T1]{fontenc}
\usepackage{orcidlink}

\newcommand{\snnt}[1]{\ensuremath{\sqrt{s_{\rm NN}} = #1 \text{\,TeV}}\xspace}

\begin{document}
%%%%%%%%%%%%%%%%%%%%%%%%%%%%%%%%%%%%%%%%%%%%%%%%%%
% These are some new commands that may be useful 
% for paper writing in general. If other newcommands
% are needed for your specific paper, please feel 
% free to add here. 
%
% The currently available commands are organized in: 
% 1) Systems
% 2) Quantities
% 3) Energies and units
% 4) Detectors
% 5) particle species 
%%%%%%%%%%%%%%%%%%%%%%%%%%%%%%%%%%%%%%%%%%%%%%%%%%

% 1) SYSTEMS 
\newcommand{\pp}           {pp\xspace}
\newcommand{\ppbar}        {\mbox{$\mathrm {p\overline{p}}$}\xspace}
\newcommand{\XeXe}         {\mbox{Xe--Xe}\xspace}
\newcommand{\PbPb}         {\mbox{Pb--Pb}\xspace}
\newcommand{\pA}           {\mbox{pA}\xspace}
\newcommand{\pPb}          {\mbox{p--Pb}\xspace}
\newcommand{\AuAu}         {\mbox{Au--Au}\xspace}
\newcommand{\dAu}          {\mbox{d--Au}\xspace}

% 2) QUANTITIES 
\newcommand{\s}            {\ensuremath{\sqrt{s}}\xspace}
\newcommand{\snn}          {\ensuremath{\sqrt{s_{\mathrm{NN}}}}\xspace}
\newcommand{\pt}           {\ensuremath{p_{\rm T}}\xspace}
\newcommand{\meanpt}       {$\langle p_{\mathrm{T}}\rangle$\xspace}
\newcommand{\ycms}         {\ensuremath{y_{\rm CMS}}\xspace}
\newcommand{\ylab}         {\ensuremath{y_{\rm lab}}\xspace}
\newcommand{\etarange}[1]  {\mbox{$\left | \eta \right |~<~#1$}}
\newcommand{\yrange}[1]    {\mbox{$\left | y \right |~<~#1$}}
\newcommand{\dndy}         {\ensuremath{\mathrm{d}N_\mathrm{ch}/\mathrm{d}y}\xspace}
\newcommand{\dndeta}       {\ensuremath{\mathrm{d}N_\mathrm{ch}/\mathrm{d}\eta}\xspace}
\newcommand{\avdndeta}     {\ensuremath{\langle\dndeta\rangle}\xspace}
\newcommand{\dNdy}         {\ensuremath{\mathrm{d}N_\mathrm{ch}/\mathrm{d}y}\xspace}
\newcommand{\Npart}        {\ensuremath{N_\mathrm{part}}\xspace}
\newcommand{\Ncoll}        {\ensuremath{N_\mathrm{coll}}\xspace}
\newcommand{\dEdx}         {\ensuremath{\textrm{d}E/\textrm{d}x}\xspace}
\newcommand{\RpPb}         {\ensuremath{R_{\rm pPb}}\xspace}

% 3) ENERGIES, UNITS
\newcommand{\nineH}        {$\sqrt{s}~=~0.9$~Te\kern-.1emV\xspace}
\newcommand{\seven}        {$\sqrt{s}~=~7$~Te\kern-.1emV\xspace}
\newcommand{\twoH}         {$\sqrt{s}~=~0.2$~Te\kern-.1emV\xspace}
\newcommand{\twosevensix}  {$\sqrt{s}~=~2.76$~Te\kern-.1emV\xspace}
\newcommand{\five}         {$\sqrt{s}~=~5.02$~Te\kern-.1emV\xspace}
\newcommand{\twosevensixnn}{$\sqrt{s_{\mathrm{NN}}}~=~2.76$~Te\kern-.1emV\xspace}
\newcommand{\fivenn}       {$\sqrt{s_{\mathrm{NN}}}~=~5.02$~Te\kern-.1emV\xspace}
\newcommand{\fiveff}       {$\sqrt{s_{\mathrm{NN}}}~=~5.44$~Te\kern-.1emV\xspace}
\newcommand{\LT}           {L{\'e}vy-Tsallis\xspace}
\newcommand{\GeVc}         {Ge\kern-.1emV/$c$\xspace}
\newcommand{\MeVc}         {Me\kern-.1emV/$c$\xspace}
\newcommand{\TeV}          {Te\kern-.1emV\xspace}
\newcommand{\GeV}          {Ge\kern-.1emV\xspace}
\newcommand{\MeV}          {Me\kern-.1emV\xspace}
\newcommand{\GeVmass}      {Ge\kern-.2emV/$c^2$\xspace}
\newcommand{\MeVmass}      {Me\kern-.2emV/$c^2$\xspace}
\newcommand{\lumi}         {\ensuremath{\mathcal{L}}\xspace}

% 4) DETECTORS 
\newcommand{\ITS}          {\rm{ITS}\xspace}
\newcommand{\TOF}          {\rm{TOF}\xspace}
\newcommand{\ZDC}          {\rm{ZDC}\xspace}
\newcommand{\ZDCs}         {\rm{ZDCs}\xspace}
\newcommand{\ZNA}          {\rm{ZNA}\xspace}
\newcommand{\ZNC}          {\rm{ZNC}\xspace}
\newcommand{\SPD}          {\rm{SPD}\xspace}
\newcommand{\SDD}          {\rm{SDD}\xspace}
\newcommand{\SSD}          {\rm{SSD}\xspace}
\newcommand{\TPC}          {\rm{TPC}\xspace}
\newcommand{\TRD}          {\rm{TRD}\xspace}
\newcommand{\VZERO}        {\rm{V0}\xspace}
\newcommand{\VZEROA}       {\rm{V0A}\xspace}
\newcommand{\VZEROC}       {\rm{V0C}\xspace}
\newcommand{\Vdecay} 	   {\ensuremath{V^{0}}\xspace}

% 4) PARTICLE SPECIES 
\newcommand{\ee}           {\ensuremath{e^{+}e^{-}}} 
\newcommand{\pip}          {\ensuremath{\pi^{+}}\xspace}
\newcommand{\pim}          {\ensuremath{\pi^{-}}\xspace}
\newcommand{\kap}          {\ensuremath{\rm{K}^{+}}\xspace}
\newcommand{\kam}          {\ensuremath{\rm{K}^{-}}\xspace}
\newcommand{\pbar}         {\ensuremath{\rm\overline{p}}\xspace}
\newcommand{\kzero}        {\ensuremath{{\rm K}^{0}_{\rm{S}}}\xspace}
\newcommand{\lmb}          {\ensuremath{\Lambda}\xspace}
\newcommand{\almb}         {\ensuremath{\overline{\Lambda}}\xspace}
\newcommand{\Om}           {\ensuremath{\Omega^-}\xspace}
\newcommand{\Mo}           {\ensuremath{\overline{\Omega}^+}\xspace}
\newcommand{\X}            {\ensuremath{\Xi^-}\xspace}
\newcommand{\Ix}           {\ensuremath{\overline{\Xi}^+}\xspace}
\newcommand{\Xis}          {\ensuremath{\Xi^{\pm}}\xspace}
\newcommand{\Oms}          {\ensuremath{\Omega^{\pm}}\xspace}
\newcommand{\degree}       {\ensuremath{^{\rm o}}\xspace}

% GENERAL
\newcommand{\la}{\langle}
\newcommand{\ra}{\rangle}
\newcommand{ \mean }[1]{\la #1 \ra}   

% FLOW
\newcommand{ \Num }[2]{\mathrm{N}\mean{#1}_{#2}}
\newcommand{ \Den }[2]{\mathrm{D}\mean{#1}_{#2}}

%%%%%%%%%%%%%%%  Title page %%%%%%%%%%%%%%%%%%%%%%%%
\begin{titlepage}
% the dates below correspond to CERN approval
% please don't touch: EB chairs will take care
\PHyear{2026}       % required, will be obtained from CERN
\PHnumber{296}      % required, will be obtained from CERN
\PHdate{03 November}  % required, will be obtained from CERN
%%%%%%%%%%%%%%%%%%%%%%%%%%%%%%%%%%%%%%%%%%%%%%%%%%%%

%%% Put your own title + short title here:
\title{Correlations and fluctuations of flow with event shape engineering in Pb--Pb and Xe--Xe collisions at the LHC}
\ShortTitle{Correlation and fluctuation of flow with ESE}   % appears on left page headers

%%% Do not change the next lines
\Collaboration{ALICE Collaboration\thanks{See Appendix~\ref{app:collab} for the list of collaboration members}}
\ShortAuthor{ALICE Collaboration} % appears on right page headers, do not change

\begin{abstract}
Measurements of anisotropic flow provide some of the strongest experimental constraints on the collective expansion and transport properties of the quark--gluon plasma produced in ultra-relativistic heavy-ion collisions. In this paper, correlations and fluctuations of the anisotropic flow coefficients $v_{2}$, $v_{3}$, and $v_{4}$ are measured in Pb--Pb and Xe--Xe collisions at $\sqrt{s_{\mathrm{NN}}}=5.02$ and $5.44$~TeV, respectively, with the ALICE detector. The Event Shape Engineering technique, based on the reduced flow vector $q_{n}$, is used to study flow correlations for different initial-state geometries. A strong anticorrelation between $v_2$ and $v_3$ and a positive correlation between $v_2$ and $v_4$ are observed. A small anticorrelation is seen between $v_3$ and $v_4$. From the $v_2$--$v_4$ correlation, the linear and nonlinear components of $v_4$ are extracted and compared to those extracted using multiparticle correlations.
The linear and nonlinear components of $v_{4}$ are also measured directly as a function of $q_2$ for the first time in Pb--Pb and Xe--Xe, revealing a strong coupling of the nonlinear term to the initial-state ellipticity $\varepsilon_2$, while the linear component shows little dependence on initial-state geometry. Additionally, the coupling coefficient $\chi_{4,22}$, which measures the nonlinear contribution of $v_2^2$ to $v_4$, is extracted and shows little dependence on $q_2$, except in very central collisions. The effects of flow fluctuations for constrained initial-state geometries are studied for the first time using Event Shape Engineering with the ratio of $v_2\{2\}$ and $v_2\{4\}$; the results show that collisions with the lowest $q_2$ exhibit more significant $v_2$ fluctuations compared to the unbiased ratio. The highest–$q_2$ events are found to have reduced flow fluctuations, corresponding to a smaller event–by–event variation of the elliptic anisotropy.
\end{abstract}
\end{titlepage}

\setcounter{page}{2} %please do not remove this line

%%%%%%%%%%%%%%%%%%%%%%%%%%%%%%%%
% begin main text
%%%%%%%%%%%%%%%%%%%%%%%%%%%%%%%%
\section{Introduction}
Experiments at the Large Hadron Collider (LHC) at CERN and the Relativistic Heavy Ion Collider (RHIC) at Brookhaven National Laboratory have played a pivotal role in advancing our understanding
of quantum chromodynamics (QCD), and in particular the quark--gluon plasma
(QGP)~\cite{Shuryak:1978ij,Shuryak:1980tp,ALICE:2022wpn,CMS:2024krd,STAR:2005gfr,PHENIX:2004vcz,BRAHMS:2004adc,PHOBOS:2004zne}.
The QGP, a state of matter that might have existed shortly after the Big Bang, is formed in high-energy heavy-ion collisions at the LHC and RHIC. The space--time evolution of
the QGP can be described by relativistic viscous hydrodynamic
models~\cite{Gale:2013da,Heinz:2013th,Song:2017wtw} with a very low shear viscosity to entropy density ratio, indicating an extremely strongly interacting nearly-perfect fluid~\cite{Kovtun:2004de}.
As a nearly-perfect fluid exhibits almost no diffusion or dissipation, spatial
correlations of the initial state are almost reversibly mapped onto the final-state momentum distribution, a phenomenon known as anisotropic flow~\cite{Ollitrault:1992bk,Voloshin:1994mz}. Therefore, the initial state can be \emph{imaged} through measurements of the momentum anisotropies of final-state particles~\cite{E877:1996czs,Poskanzer:1998yz}. 
These anisotropies originate not only from the average geometric overlap of the colliding nuclei but also from event-by-event fluctuations in the spatial distribution of participating nucleons, which generate higher-order spatial anisotropies and contribute significantly to the observed flow patterns~\cite{PhysRevC.81.054905}.

The anisotropic flow of particles is typically described by coefficients, $v_n$, of a Fourier decomposition of the azimuthal distribution of produced particles, 
\begin{align}
	\frac{d{\rm N}}{d\varphi} \propto 1 + 2 \sum_{n=1}^{\infty} v_n e^{i{n(\varphi-\Psi_n)}}, \label{eq:eq1}
\end{align}
where $\varphi$ is the azimuthal angle and $\Psi_n$ is the angle of the $n^{\rm th}$-order symmetry plane.  Measurements of the $v_n$ coefficients provide significant insight into the thermodynamic properties of the QGP produced in heavy-ion collisions~\cite{PHENIX:2006dpn,STAR:2022ncy,
ALICE:2010suc,ALICE:2011ab,ALICE:2016ccg,ALICE:2018lao,ATLAS:2012at,ATLAS:2019dct,CMS:2012tqw,CMS:2019cyz,LHCb:2023png}.

This paper concerns the elliptic flow, $v_2$, the
triangular flow, $v_3$, the quadrangular flow, $v_4$, and the correlations between them. If the
initial state is characterized by the spatial anisotropies
$\varepsilon_n$, where $\varepsilon_2$ quantifies its \textit{ellipticity} while $\varepsilon_3$ reflects its \textit{triangularity}, then $v_2 \propto \varepsilon_2$ and $v_3 \propto
\varepsilon_3$~\cite{Niemi:2012aj}. However, $v_4$ comprises both a ``linear'' contribution associated with the cumulant-defined eccentricity $\epsilon_4'$~\cite{Teaney_2014,Qian:2017ier}, which subtracts contributions from lower-order moments as introduced in Eq.~(2.3) of Ref.~\cite{Teaney_2014}, and a ``nonlinear'' component that depends on $\varepsilon_2^2$~\cite{Yan:2015jma,Qian:2016fpi}. Disentangling these linear and nonlinear contributions provides additional insights into the QGP's hydrodynamic response, in particular $\eta/s$, to initial conditions~\cite{ALICE:2017fcd,ALICE:2020sup}. It also provides sensitivity to event–by–event fluctuations in the spatial distribution of the participating nucleons that affect the flow harmonics and correlations between different symmetry planes generated by nonlinear mode couplings that develop during the hydrodynamic evolution of the medium~\cite{Yan:2015jma,Giacalone:2016afq,ATLAS:2014ndd}. 

The higher harmonics are particularly sensitive to $\eta/s$~\cite{ALICE:2011ab} and, for the harmonics $n>3$, they are more complex, with terms arising from both $\varepsilon_n$ and nonlinear mixing of lower harmonics~\cite{Teaney_2014}. The anisotropic flow is commonly expressed in terms of the complex flow vectors $V_n = v_n e^{in\Psi_n}$. In particular, the fourth–order flow vector can be decomposed as
\begin{align}
    V_4 = V_4^\mathrm{L} + V_4^\mathrm{NL} = V_4^\mathrm{L} + \chi_{4,22} (V_2)^2. 
    \label{eq:v4parts}
\end{align} 
Here, $V_4^\mathrm{L}$ denotes the linear contribution associated with the fourth-order eccentricity $\epsilon_4'$ defined above by subtracting contributions from lower-order moments, which is uncorrelated with $\varepsilon_2^2$, while the second term represents the nonlinear response generated by mode coupling of the elliptic flow $V_2$. The coefficient $\chi_{4,22}$ quantifies the strength of the nonlinear response and has been previously measured in Pb--Pb collisions~\cite{ALICE:2017fcd,ALICE:2020sup}. More information about the hydrodynamic evolution of the QGP can therefore be obtained through detailed studies of the correlations between the flow harmonics. Symmetric cumulants, denoted as SC($n, m$), quantify the strength of the correlation between different flow harmonics $v_n^2$ and $v_m^2$~\cite{Bilandzic:2013kga}. The specific cumulants SC($4, 2$) and SC($3, 2$) describe the correlation between $v_2^2$ and $v_4^2$, and $v_2^2$ and $v_3^2$, respectively. Previous measurements in Pb--Pb collisions show positive correlations in SC($4, 2$) and anticorrelations in SC($3, 2$)~\cite{ALICE:2016kpq, ALICE:2021adw}. The positive SC($4,2$) indicates that events with larger $v_2$ tend to also have larger $v_4$, which is expected from the nonlinear contribution to $v_4$ driven by $v_2^2$~\cite{Giacalone:2016afq}. In contrast, the negative SC($3,2$) measures an anticorrelation between $v_2^2$ and $v_3^2$, consistent with the largely fluctuation-driven origin of $v_3$ compared to the elliptic geometry driving $v_2$~\cite{Zhu:2016puf}. The measurements of these symmetric cumulants have established characteristic correlation patterns between flow harmonics, providing a baseline for the more differential approach used in this paper.

In addition to fluctuations in the spatial distribution of participating nucleons, the intrinsic shape of the colliding nuclei can influence the initial–state geometry. While $^{208}$Pb is approximately spherical, $^{129}$Xe exhibits a quadrupole deformation~\cite{RevModPhys.75.121}. In particular, nuclear deformation influences the development of anisotropic flow, especially in the most central collisions, where it contributes to the initial-state eccentricity and leads to an enhanced elliptic flow in Xe–Xe collisions compared to Pb–Pb collisions~\cite{Bally_2022,ALICE:2018lao,atlasXe,CMS:2019cyz}.

Event Shape Engineering (ESE)~\cite{Schukraft:2012ah,ALICE:2015lib} has emerged as a powerful tool for studying the impact of the initial-state geometry on a variety of observables. The ESE technique consists of selecting events based on the magnitude of the final-state reduced flow vector $q_n$. Although $q_n$ can also be affected by other sources of anisotropy, the use of a rapidity separation between the event-shape selection and the measured flow harmonics suppresses non-flow effects and increases the sensitivity to the asymmetry of the initial-state geometry~\cite{STAR:2002hbo,ATLAS:2015qwl}. By examining the dependence of flow harmonics on $q_n$, one can probe the coupling strength between the linear and nonlinear flow modes. For example, since the nonlinear contributions to $v_4$ are proportional to $\varepsilon_2^2$, they are expected to vary significantly with $q_2$, while the linear contributions ($\propto \varepsilon_4$) should remain largely independent of the initial ellipticity. This has been demonstrated in the AMPT transport model, which simulates the dynamical evolution of heavy-ion collisions~\cite{Magdy:2020bhd}, but not yet in experimental data. Furthermore, the nonlinear response coefficient $\chi_{4,22}$ can be directly extracted from the nonlinear component of $v_4$ under the assumption that it is independent of the initial-state ellipticity $\varepsilon_2$~\cite{ALICE:2024nqd}. In this paper, we also test this assumption by studying how $\chi_{4,22}$ couples to $q_2$.

This paper presents a study of the correlation between different flow
coefficients in Pb--Pb and Xe--Xe collisions at \snnt{5.02} and \snnt{5.44}, respectively, which makes it possible to investigate the system size dependence. The correlations between different flow harmonics, e.g., $v_2$--$v_3$, $v_2$--$v_4$, are investigated in the midrapidity region for events selected according to $q_2$ measured at forward rapidity. The linear and nonlinear contributions
to higher-order flow coefficients are explored, and ESE is employed to investigate the dependence of $v_4$ on the event shape
and to provide a detailed analysis of the coupling strength between the different
flow contributions.

\section{Observables} 

The flow harmonics, $v_n$, can be determined by the correlation between produced particles and the symmetry plane, 
\begin{align}
	\langle v_n \rangle = \langle \langle  e^{in( \varphi - \Psi_n )} \rangle \rangle,
\end{align}
where the double brackets indicate first an average over particles in each event, and then an average over all events. Alternatively, to avoid the necessity of calculating $\Psi_n$ in each event, $v_n$ can be measured from multiparticle correlations.

The flow coefficients from two-particle correlations are defined through the two-particle cumulant, $c_n\{2\}$, as
\begin{align}
	v_n\{2\} \equiv c_n\{2\}^{1/2} =  \langle \langle e^{in ( \varphi_1 - \varphi_2 )} \rangle \rangle^{1/2}. \label{eq:2part}
\end{align}
These can be sensitive to non-flow effects, such as jets or resonance decays, that give rise to azimuthal correlations unrelated to the initial geometry. To suppress non-flow effects, the tracks used for the $v_n$ measurement in a two-particle correlation are selected from two sub-events separated by a pseudorapidity gap.

Flow involves correlations among many particles in an event, whereas non-flow effects typically involve only a few particles. The contribution from non-flow effects can therefore be reduced by evaluating the four-particle cumulant
\begin{align}
    c_n\{4\} &= \langle \langle e^{in(\phi_1 + \phi_2 - \phi_3 - \phi_4)}\rangle\rangle -2\cdot\langle \langle e^{in(\phi_1- \phi_2)}\rangle\rangle^2, \\
    v_n\{4\} &= \sqrt[4]{-c_n\{4\}}, 
\end{align}
where contributions from two-particle correlations are explicitly subtracted. 

Equation~\ref{eq:eq1} leads naturally to a geometric description of the medium in terms of a set of vectors, 
\begin{equation}
Q_n = \sum_{i=1}^{N} w_i e^{in\varphi_i},
\label{eq:Qvec}
\end{equation}
where the summation runs over the produced particles or energy distribution of the event, and the corresponding weights $w_i$ are used to correct for the non-uniform detector acceptance. 
The direction of each $Q_n$ vector is the angle of the $n^{\rm th}$-order symmetry plane, $\Psi_n$, and the magnitude is related to $v_n$.

To focus more specifically on the shape of the collision, the inherent multiplicity dependence in the magnitude of $Q_n$ is factored out to obtain the reduced flow vector
\begin{align}
	q_n = \frac{|\pmb{Q}_n|}{\sqrt{M}}, 
    \label{eq:Qvec_red}
\end{align}
where $M = \sum_{i=1}^{N} w_i$.  

The reduced flow vector $q_n$ can be measured in a particular region of phase space, i.e., at forward rapidity, which allows for a classification of events according to the flow in that region. This flow is expected to be dominated by the geometry of the initial state, but can also include contributions from other sources of anisotropy. 

Within a narrow centrality range, and by varying the selection on $q_n$, physics effects such as flow fluctuations and correlations in other regions of phase space can be studied multi-differentially. 

\subsection{Correlation between $v_2$ and $v_3$ with two-particle cumulants}

To quantify the correlation between $v_2\{2\}$ and $v_3\{2\}$, the data sample is subdivided into ranges of $q_2$ within each centrality interval, and $v_2\{2\}$ and $v_3\{2\}$ are determined for each range. The dependence of $v_3\{2\}$ on $v_2\{2\}$ is then fitted with a linear function
\begin{equation}
	v_3 = v_3^0 + k v_2,
	\label{eq:v3v2}
\end{equation}
where $v_3^0$ represents the intercept and $k$ is the slope. 

\subsection{Correlation between $v_2$ and $v_4$ with two-particle cumulants}

The fourth-harmonic flow vector, $V_4$, can be written as a sum of two uncorrelated components~\cite{Yan:2015jma,Qian:2016fpi} (see Eq.~\ref{eq:v4parts}).

When measured flow magnitudes are considered after event averaging, the uncorrelated linear and nonlinear contributions are added in quadrature.

To determine the unbiased values of $v_4^\mathrm{L}$ and $v_4^\mathrm{NL}$ using the $v_2$--$v_4$ correlation obtained with ESE, the correlation is fitted in each centrality interval with the functional form, 
\begin{equation}
	v_4 = \sqrt{c_0^2 + (c_1 v_2^2)^2},
	\label{eq:v4fit}
\end{equation}
where $c_0$ represents the linear contribution and $c_1$ quantifies the nonlinear dependence of $v_4$ on $v_2$. This functional form has been employed in previous studies to disentangle the various contributions to higher-order flow harmonics when using the ESE procedure, and the fitted parameters provide valuable insights into the nature of the correlation~\cite{ATLAS:2015qwl}.
From the fit results, the linear and nonlinear flow modes of $v_4$ in each centrality interval are then extracted as
\begin{equation}
	v_4^\mathrm{L} = c_0, ~v_4^\mathrm{NL} = \sqrt{ v_4^2 - c_0^2 }.
\label{eq:extractv4nl}
\end{equation}

\subsection{Extraction of $v_4^\mathrm{NL}$ and $v_4^\mathrm{L}$ using multi-particle correlations}
 
Alternatively, it is possible to directly obtain the nonlinear contribution to $v_4$ by measuring
\begin{align}
    v_{4,22}= \frac{\langle \langle e^{i( 4\phi_1 - 2\phi_2 - 2\phi_3)} \rangle\rangle}{\sqrt{\langle \langle e^{i(2\phi_1 + 2\phi_2 - 2\phi_3 - 2\phi_4)}\rangle\rangle}} \sim \langle v_4 e^{i (4\Psi_4 - 4\Psi_2)} \rangle. \label{eq:v4nl}
\end{align}

Using Eq.~\ref{eq:v4parts}, this nonlinear part is then removed from the measured $v_4^2$ in quadrature to obtain the linear response term. The linear and nonlinear response terms will serve as baseline measurements when comparing the extracted response using $v_2$--$v_4$ correlations with ESE.

The nonlinear flow mode coefficient $\chi_{4,22}$~\cite{Yan:2015jma} is extracted by dividing $v_4^{\rm NL}$ by $v_2^2$
\begin{align}
    \chi_{4,22} = \frac{v_4^\mathrm{NL}}{\sqrt{\langle\langle e^{i(2\phi_1 + 2\phi_2 - 2\phi_3 - 2\phi_4)}\rangle\rangle}}. \label{eq:chi422}
\end{align}
Comparing $v_4^\mathrm{L}$ and $v_4^\mathrm{NL}$ obtained directly from multiparticle cumulants with the components extracted from the ESE-based $v_2$--$v_4$ correlation tests whether the linear part is independent of $\varepsilon_2$ and whether the nonlinear part follows the expected coupling to $v_2^2$. Testing the $\chi_{4,22}$ coupling to $q_2$ will reveal whether it is independent of the initial elliptic eccentricity.

\subsection{Fluctuations in the underlying $v_2$ distribution}

The two- and four-particle cumulant flow coefficients, $v_n\{2\}$ and $v_n\{4\}$, reflect different moments of the underlying distribution describing the flow fluctuations of $v_n$. Assuming that the flow fluctuations follow a Bessel-Gaussian distribution in the limit of small fluctuations ($\sigma_{v_n} \ll \langle v_n \rangle$), the flow harmonics $v_n\{m\}$ will have opposite contributions from the fluctuations, as expressed by the following relations~\cite{Voloshin:2007pc,Ollitrault:2009ie}
\begin{align}
	v_n\{2\}^2 &= \langle v_n \rangle^2 + \sigma_{v_n}^2 + \delta_{v_n}^2, \label{eq:vn2} \\
	v_n\{4\}^2 &= \langle v_n \rangle^2 - \sigma_{v_n}^2. \label{eq:vn4}
\end{align}
Here, $\delta_{v_n}$ represents the contribution from non-flow effects, which are assumed to be small after imposing a $\Delta\eta$ gap between sub-events~\cite{ALICE:2022zks}. Therefore, the difference between $v_2\{2\}$ and $v_2\{4\}$ is expected to be mainly due to the fluctuations. The selection on $q_2$ may also bias the underlying flow distribution. To investigate how ESE constrains flow fluctuations, the ratio of $v_n\{4\}$ to $v_n\{2\}$ is measured in $q_2$ percentile bins. This ratio is commonly used to study flow fluctuations~\cite{pdfv2,PhysRevC.90.024903}.

\section{Data Analysis}

This study analyzes minimum bias data from Pb--Pb and Xe--Xe collisions at \fivenn and \fiveff, respectively, collected during the Run 2 period between 2015 and 2018~\cite{ALICE:2008ngc,ALICE:2014sbx}. Minimum bias events are triggered by simultaneous signals in the forward V0A ($2.8 < \eta < 5.1$) and V0C ($-3.7 < \eta < -1.7$) detectors~\cite{Cortese:781854}, which measure charged-particle production at forward rapidities. The 2018 data sample was enhanced in the 0--10\% and 30--50\% centrality ranges by event triggers based on the multiplicity in the V0A and V0C. Furthermore, only events occurring at the center of the detector (with a primary vertex reconstructed within 10 cm along the beam line from the nominal interaction point) are analyzed to ensure full tracking acceptance. Collision events are recorded with both polarities of the ALICE solenoid magnetic field.

The charged-particle tracks utilised in the measurement of the flow coefficients $v_n$ are reconstructed using the Inner Tracking System (ITS)~\cite{CERN-LHCC-99-012}, which provides precise tracking close to the interaction point, and the Time Projection Chamber (TPC)~\cite{Dellacasa:451098}, which provides the main tracking and charged-particle momentum measurements, within the pseudorapidity range \etarange{0.8} and the transverse-momentum interval $0.2 < \pt < 5.0~(3.0)~\mathrm{GeV}/c$ in Pb–Pb (Xe–Xe) collisions. Because the charged-particle spectrum falls steeply with $\pt$, particles at low $\pt$ dominate the $\pt$-integrated flow measurement~\cite{pdfv2}. Moderate variations of the upper $\pt$ limit are therefore expected to have only a small effect on the measured $v_n$. This is verified by varying the upper $\pt$ cutoff in Xe--Xe collisions, resulting in changes of about $1$--$2$\%, smaller than the systematic uncertainties and consistent with statistical fluctuations. Tracks are required to be reconstructed from at least 70 out of 159 possible TPC space points ($N_{\rm clus}$), and the track fit quality is required to be better than $\chi^2/{N_{\rm clus}} = 2.5$. Constraints on the distance of closest approach (DCA) to the primary vertex are applied. In the longitudinal direction, a restriction of $< 2~\mathrm{cm}$ is imposed. In the transverse plane, the requirement is defined as
$\sigma \cdot \left(0.0026 + 0.005/p_{\mathrm{T}}^{1.01}\right)~\mathrm{cm}$, with $\sigma = 7$ and $p_{\rm T}$ in~\GeVc. 

The anisotropic flow observables are calculated using the Generic Framework (GF)~\cite{Bilandzic:2013kga, Moravcova:2020wnf}, which is a fast and efficient algorithm for determining multiparticle correlations without nested loops. The GF incorporates corrections for the non-uniform azimuthal detector acceptance and detector efficiency.  For the two-particle correlations, particles are divided into two pseudorapidity-separated subevents, and only particle pairs formed from different subevents are correlated. Non-flow effects are suppressed by requiring a pseudorapidity separation of $|\Delta\eta|>1$ between the subevents in Pb--Pb collisions, while in Xe--Xe collisions a smaller gap of $|\Delta\eta|>0.4$ is used owing to the limited data sample size. For higher-order cumulants, no pseudorapidity separation is required between the correlated particles, as they are less sensitive to non-flow effects by definition.

The event shape, characterised by the reduced flow vector $q_n$ ($n=2,3$), is calculated from energy deposition in the V0C detector, using Eqs.~\ref{eq:Qvec} and~\ref{eq:Qvec_red}, where the sums run over all channels of the V0C ($i = 1-32$) and $w_i$ represents the energy deposition in each channel. To correct for non-uniformities in the V0C detector, a calibration procedure is conducted on a run-by-run basis, comprising two steps: gain equalization and recentering of the $\pmb{Q}_n$ vector, similar to the procedure described in Ref.~\cite{Selyuzhenkov_2008}.

Pb--Pb collision events are classified according to $q_n$ ($n=2,3$) into ten quantiles, ranging from 0--10\% to 90--100\%. The 0--10\% quantile represents the 10\% of events with the least elliptic shape, while the 90--100\% quantile corresponds to the 10\% of events with the most elliptic shape. The resolution of the V0 detector allows events to be selected in narrow 1\%-wide centrality intervals. To mitigate potential biases related to multiplicity, $q_n$ is calculated within these intervals, which are subsequently combined into broader bins. Due to the smaller size of the Xe--Xe data sample, five $q_n$ ($n=2,3$) bins are investigated, ranging from 0--20\% to 80--100\%.

A critical aspect of this analysis is reducing unwanted auto-correlations between the centrality estimation, event shape reconstruction, and flow coefficient measurement. While the centrality was estimated by percentiling the total energy deposition in the V0A and V0C detectors, the reduced flow vector $q_2$ was calculated utilising only information from the V0C detector.  The tracks involved in the measurement of $v_n$ are from the midrapidity region, resulting in a pseudorapidity separation of more than 0.9 units from the V0C.  

\subsection{Systematic Uncertainties}
The systematic uncertainties were assessed by varying individual analysis selection criteria while keeping all other selections fixed. For the event selection, the primary-vertex position along the beam line is required to be within $7$ cm of the nominal interaction point, alternative centrality estimators at midrapidity using the Silicon Pixel Detector (SPD) are employed, and results from two different magnetic field polarities are compared. For the track selection, the minimum number of TPC space points is increased up to $N_{\rm clus} \geq 90$ points, and the DCA constraints in both the longitudinal and transverse directions are tightened to $0.5$ cm and $4\sigma$, respectively. The impact of the track selection is evaluated by repeating the analysis using a modified tracking scheme which imposes different requirements on the hits in the ITS. The impact of the track filter fit quality is also checked by requiring a tighter $\chi^2/N_{\rm clus}$ selection of $2.3$. In general, the two largest contributions to the systematic uncertainty come from varying the constraint on the primary-vertex position and using different track criteria, but smaller contributions are also present from other sources. All significant contributions, determined using a Barlow test~\cite{barlow2002systematicerrorsfactsfictions}, are summed in quadrature to obtain the final systematic uncertainty. In Pb--Pb,  the total relative systematic uncertainty of $v_2$ is approximately $1$--$3\%$, for $v_3$ it is around $2$--$4\%$, and for $v_4$ it ranges from $3\%$ to $10\%$ across the different ESE quantiles. In Xe--Xe, the total systematic uncertainty across ESE quantiles is around $1$--$2\%$ for $v_2$, for $v_3$ it is $2$--$6\%$, and for $v_4$ it is $5$--$11\%$, with the largest uncertainties in peripheral collisions. The systematic uncertainty is similar for the two different collision systems, but increases for Xe--Xe for higher harmonics.

\section{Results}
\begin{figure}[H]
	\centering
	\begin{minipage}{.49\textwidth}
		\centering
        \includegraphics[width=\textwidth]{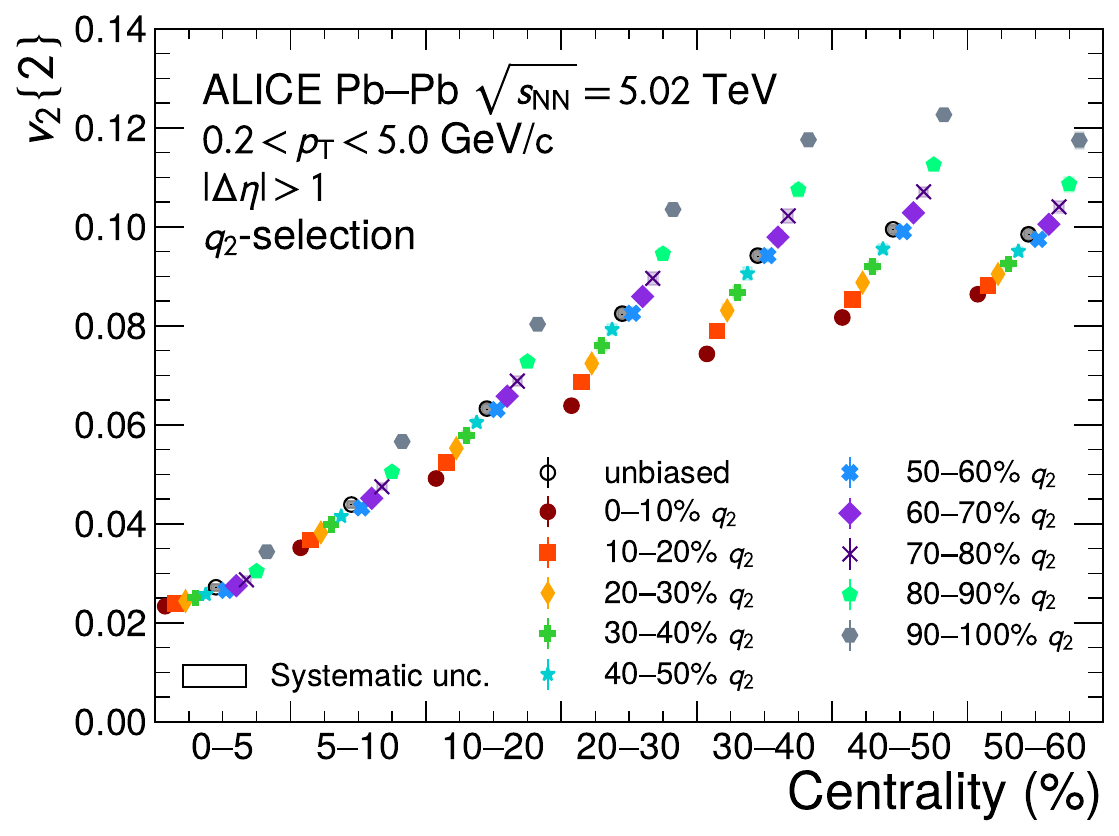}
	\end{minipage}
	\begin{minipage}{.49\textwidth}
		\centering
		\includegraphics[width=\textwidth]{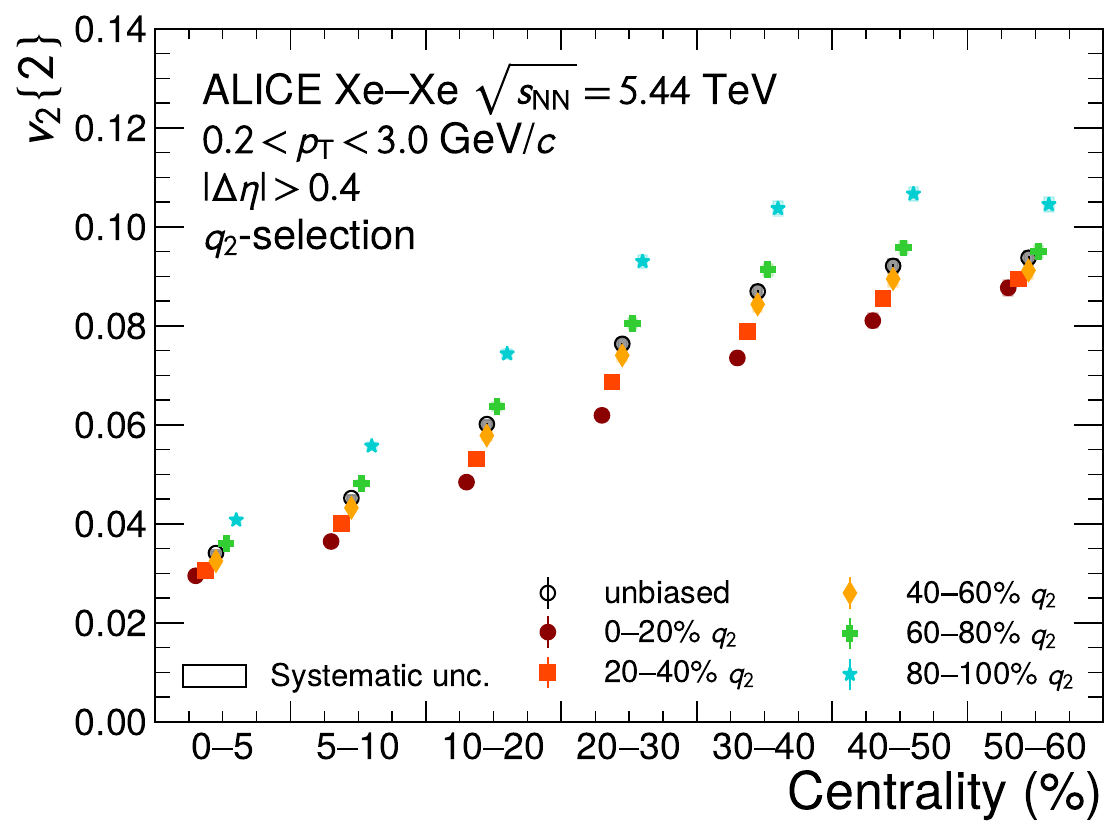}
	\end{minipage}
	\caption{Charged-particle $v_2\{2\}$ for event-shape-selected and unbiased events as a function of centrality in Pb--Pb (left) and Xe--Xe (right) collisions. Events are classified according to $q_2$ measured in the V0C detector. Points corresponding to different $q_2$ values are slightly shifted horizontally within each centrality interval for better visibility. Error bars (boxes) indicate statistical (systematic) uncertainties.}
	\label{fig:v22centPbXe}
\end{figure}

\begin{figure}[H]
	\centering
	\begin{minipage}{.49\textwidth}
		\centering
		\includegraphics[width=\textwidth]{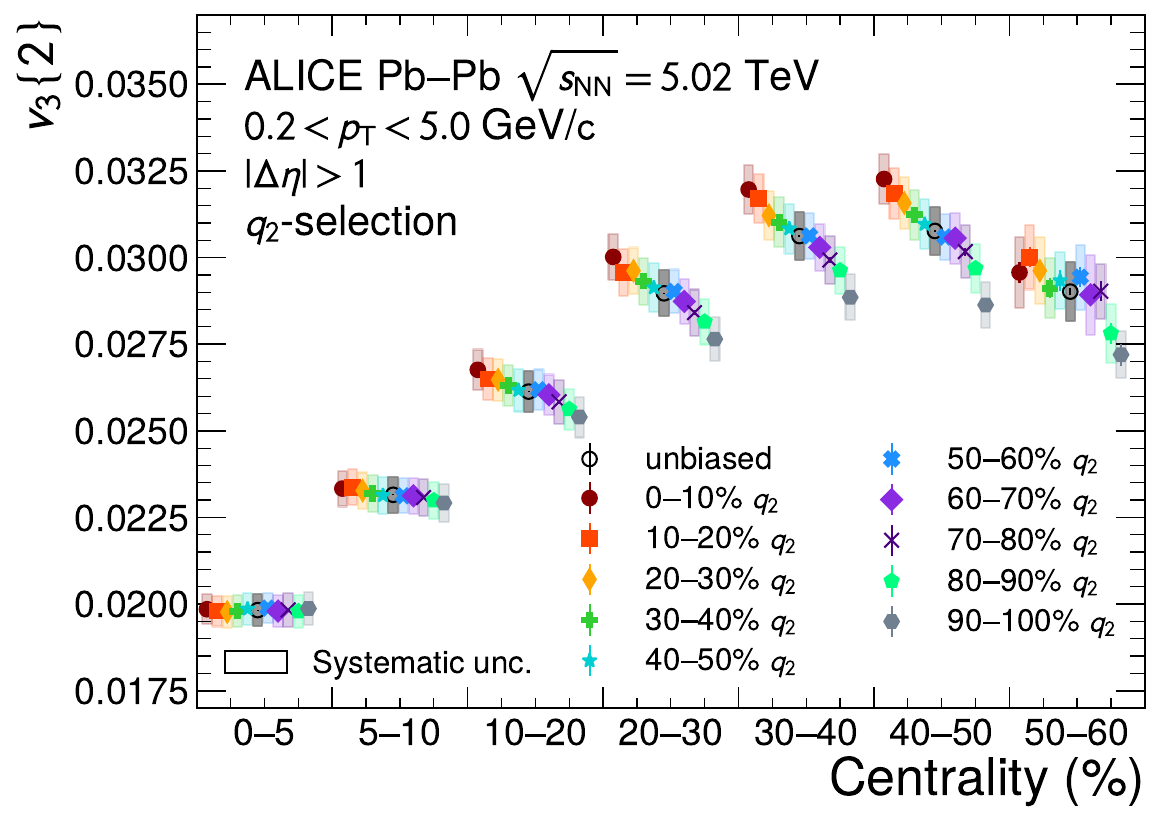}
	\end{minipage}
	\begin{minipage}{.49\textwidth}
		\centering
		\includegraphics[width=\textwidth]{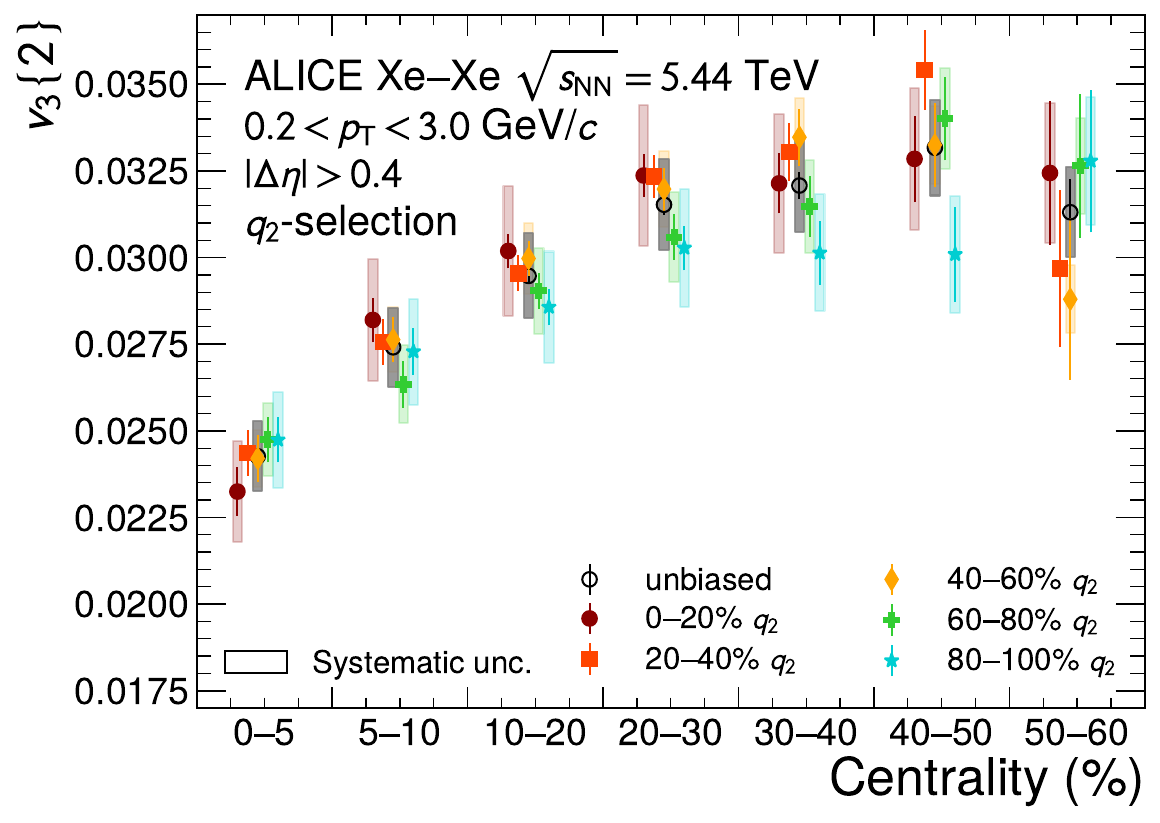}
	\end{minipage}
	\caption{Charged-particle $v_3\{2\}$ for event-shape-selected and unbiased events as a function of centrality in Pb--Pb (left) and Xe--Xe (right) collisions. Events are classified according to $q_2$ measured in the V0C detector. Points corresponding to different $q_2$ values are slightly shifted horizontally within each centrality interval for better visibility. Error bars (boxes) indicate statistical (systematic) uncertainties.}
	\label{fig:v32centPbXe}
\end{figure}

\begin{figure}[H]
	\centering
	\begin{minipage}{.49\textwidth}
		\centering
		\includegraphics[width=\textwidth]{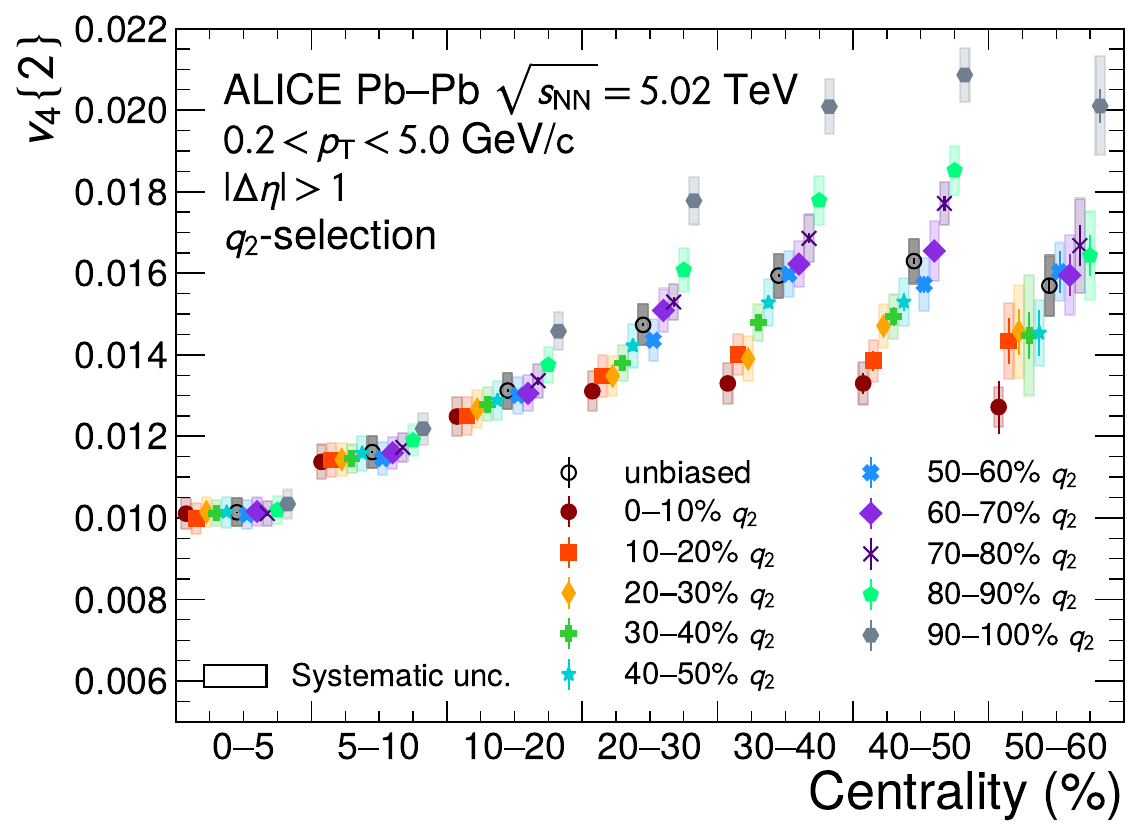}
	\end{minipage}
	\begin{minipage}{.49\textwidth}
		\centering
		\includegraphics[width=\textwidth]{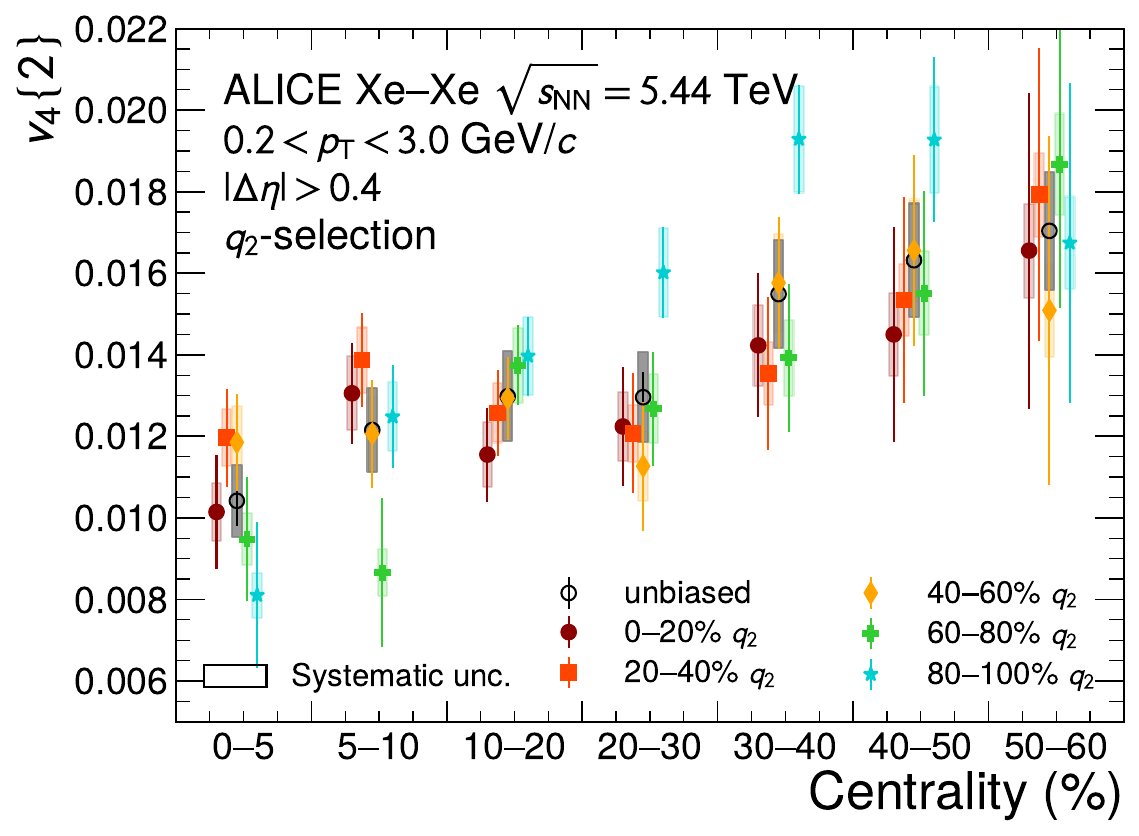}
	\end{minipage}
	\caption{Charged-particle $v_4\{2\}$ for event-shape-selected and unbiased events as a function of centrality in Pb--Pb (left) and Xe--Xe (right) collisions. Events are classified according to $q_2$ measured in the V0C detector. Points corresponding to $q_2$ values are slightly shifted horizontally within each centrality interval for better visibility. Error bars (boxes) indicate statistical (systematic) uncertainties.}
	\label{fig:v42centPbXe}
\end{figure}

\begin{figure}[H]
	\centering
	\begin{minipage}{.49\textwidth}
		\centering
		\includegraphics[width=\textwidth]{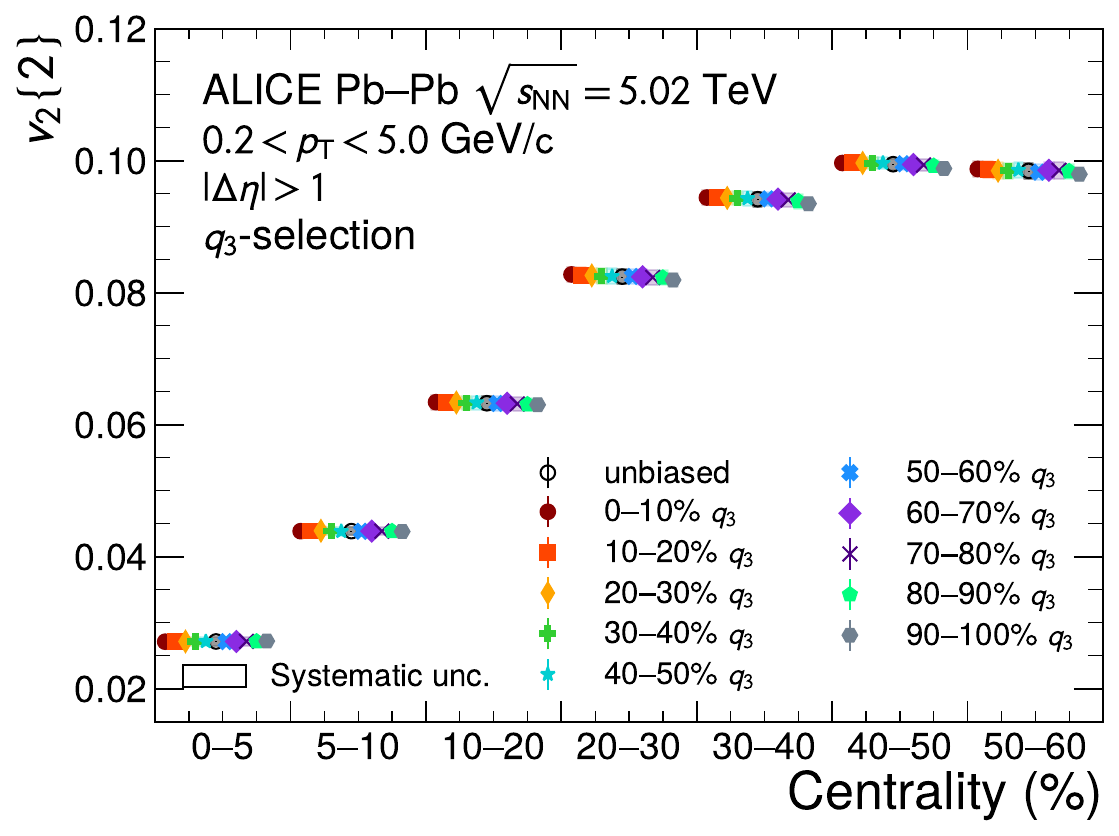}
	\end{minipage}
	\begin{minipage}{.49\textwidth}
		\centering
		\includegraphics[width=\textwidth]{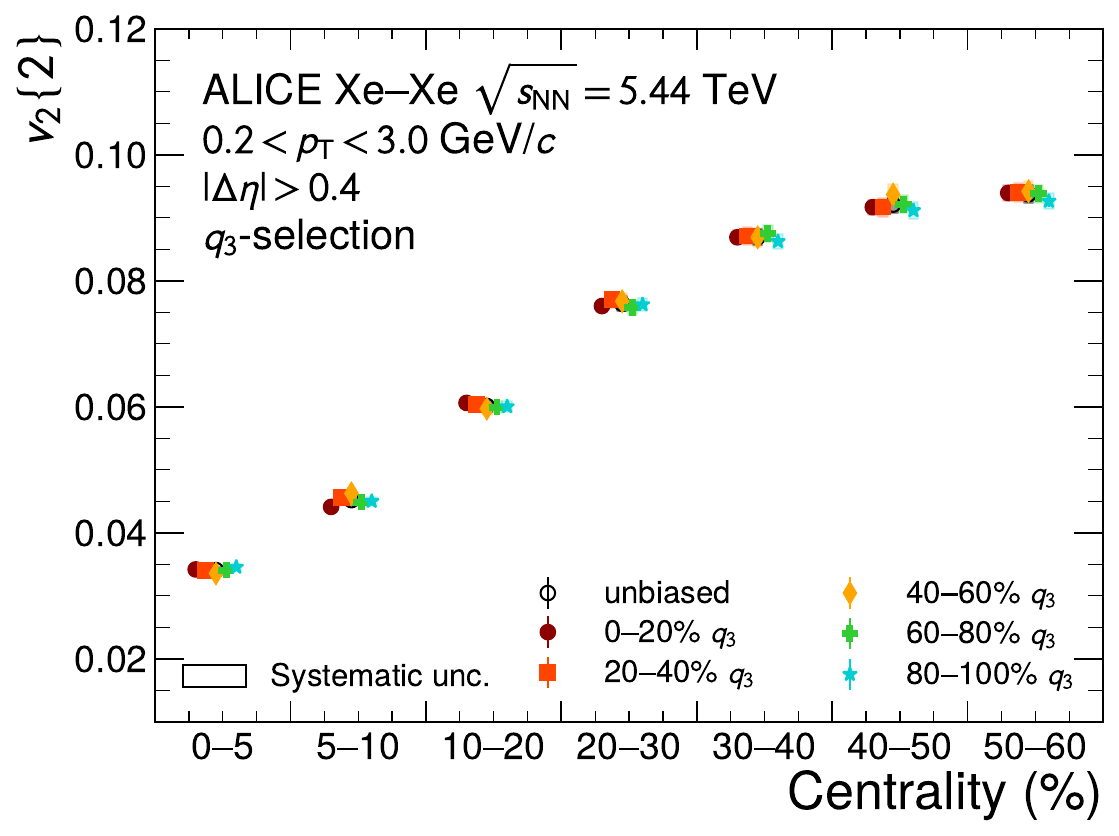}
	\end{minipage}
	\caption{Charged-particle $v_2\{2\}$ for event-shape-selected and unbiased events as a function of centrality in Pb--Pb (left) and Xe--Xe (right) collisions. Events are classified according to $q_3$ measured in the V0C detector. Points corresponding to $q_3$ values are slightly shifted horizontally within each centrality interval for better visibility. Error bars (boxes) indicate statistical (systematic) uncertainties.}
	\label{fig:v22centPbXe_q3}
\end{figure}

\begin{figure}[H]
	\centering
	\begin{minipage}{.49\textwidth}
		\centering
		\includegraphics[width=\textwidth]{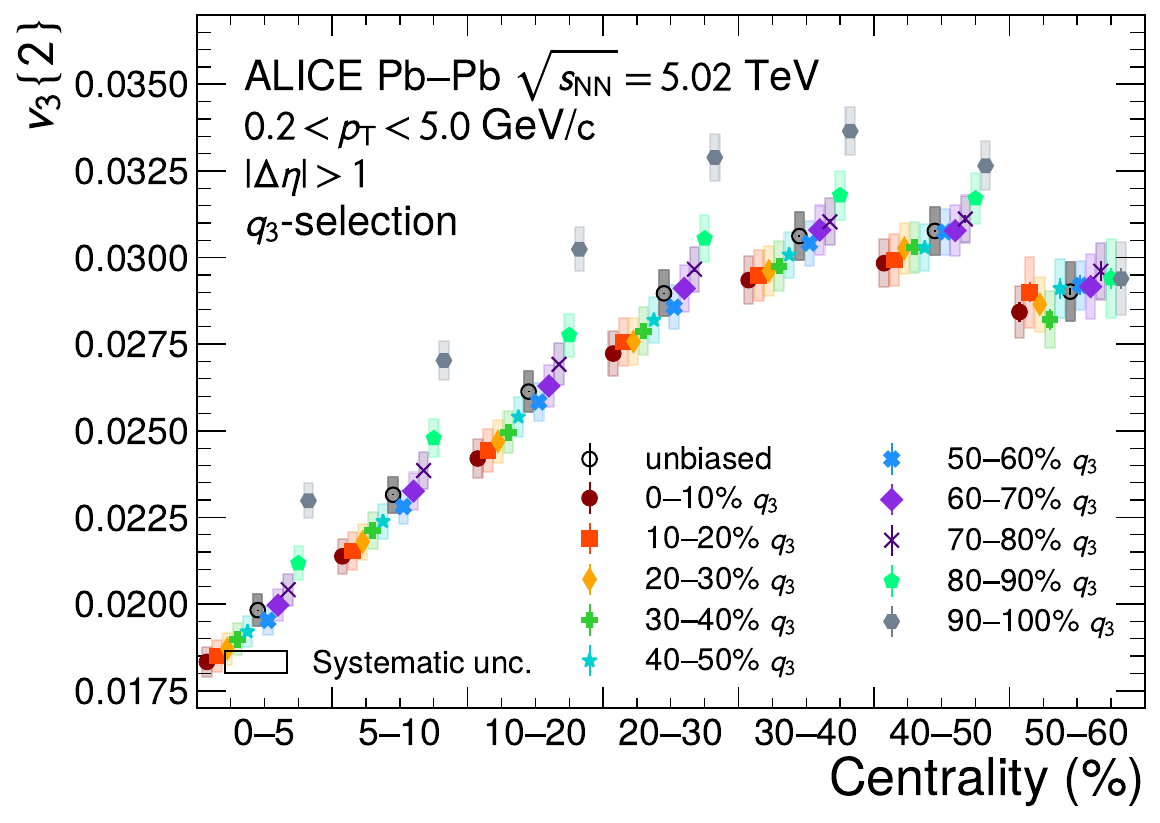}
	\end{minipage}
	\begin{minipage}{.49\textwidth}
		\centering
		\includegraphics[width=\textwidth]{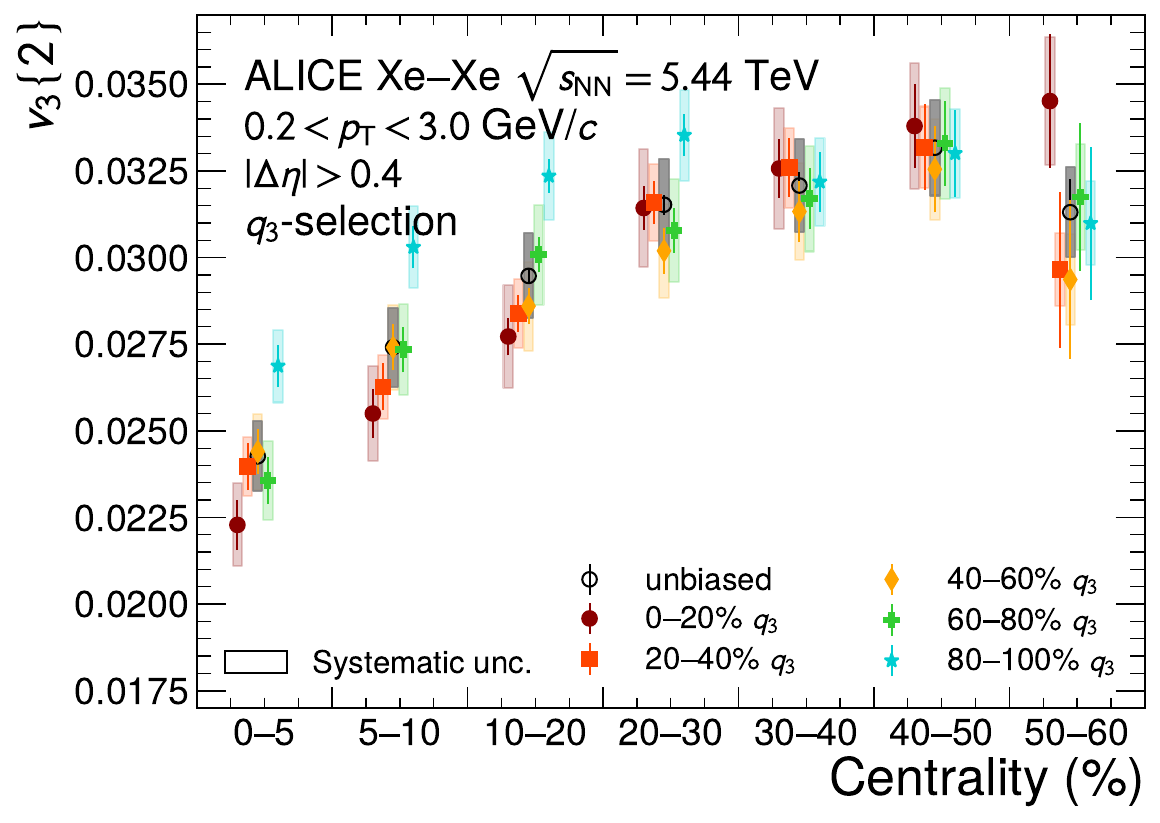}
	\end{minipage}
	\caption{Charged-particle $v_3\{2\}$ for event-shape-selected and unbiased events as a function of centrality in Pb--Pb (left) and Xe--Xe (right) collisions. Events are classified according to $q_3$ measured in the V0C detector. Points corresponding to $q_3$ values are slightly shifted horizontally within each centrality interval for better visibility. Error bars (boxes) indicate statistical (systematic) uncertainties.}
	\label{fig:v32centPbXe_q3}
\end{figure}

\begin{figure}[H]
	\centering
	\begin{minipage}{.49\textwidth}
		\centering
		\includegraphics[width=\textwidth]{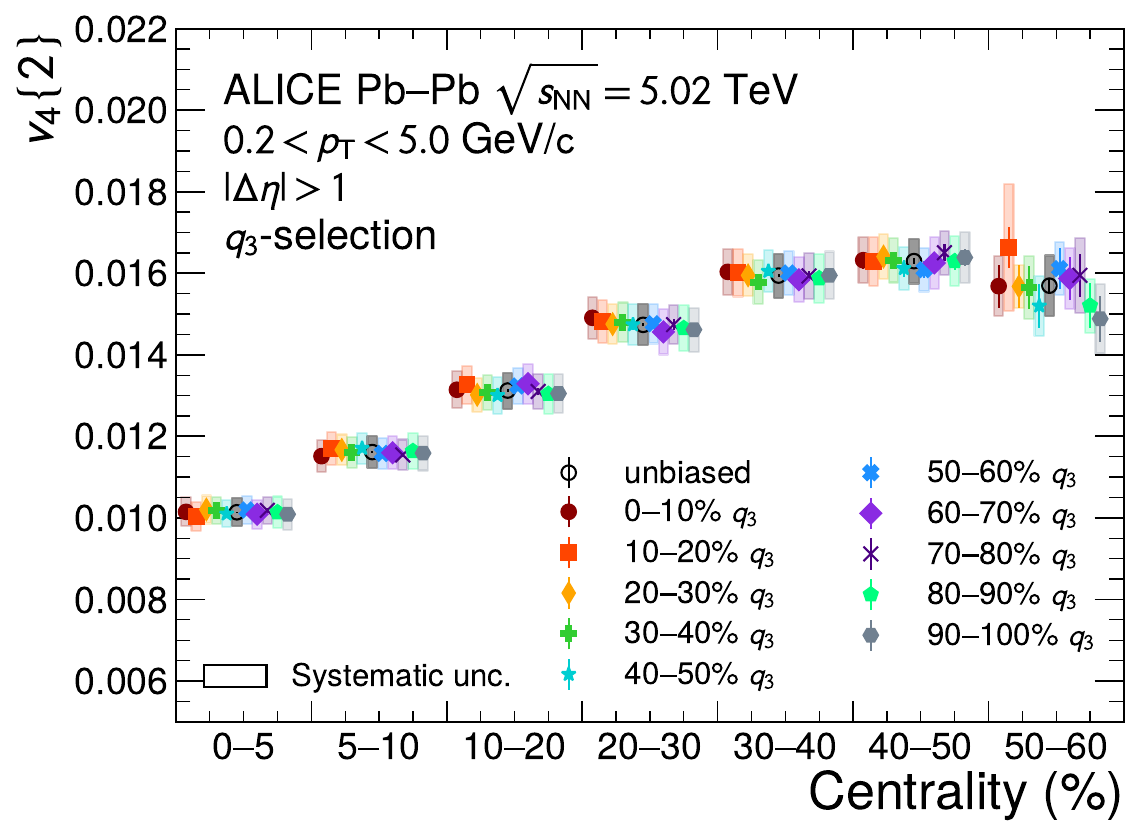}
	\end{minipage}
	\begin{minipage}{.49\textwidth}
		\centering
		\includegraphics[width=\textwidth]{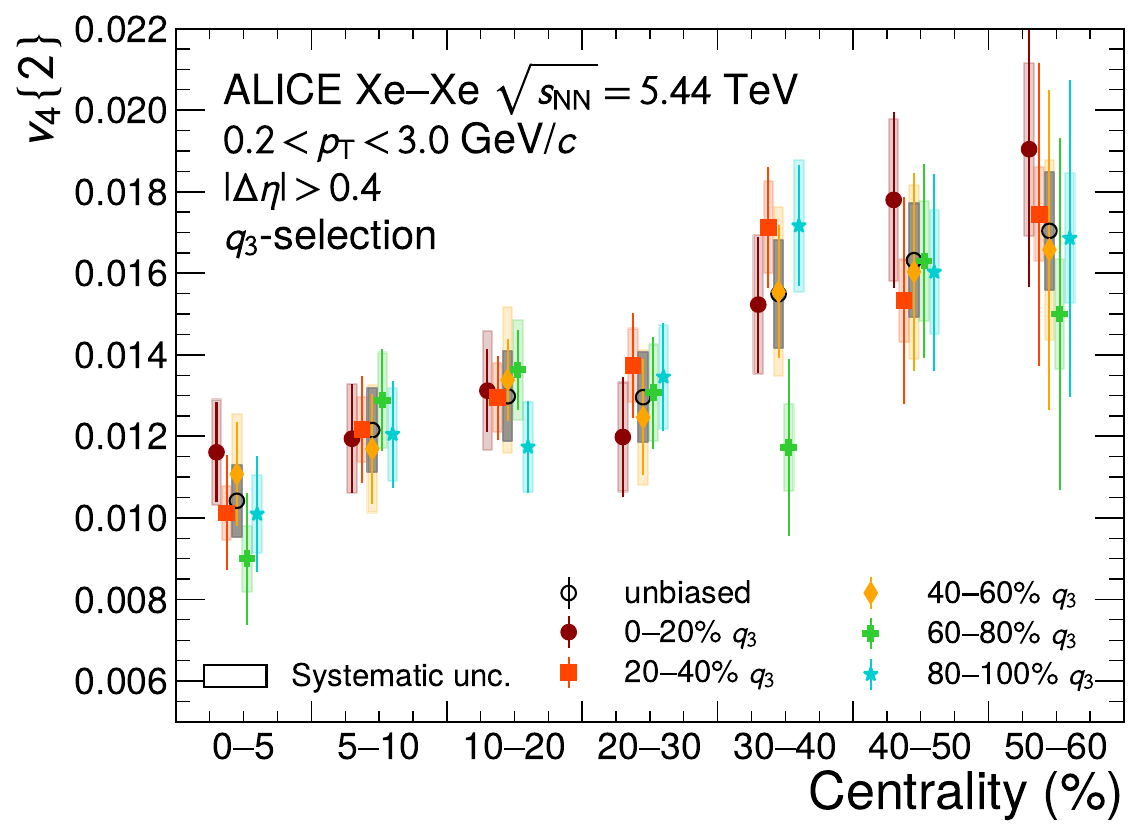}
	\end{minipage}
	\caption{Charged-particle $v_4\{2\}$ for event-shape-selected and unbiased events as a function of centrality in Pb--Pb (left) and Xe--Xe (right) collisions. Events are classified according to $q_3$ measured in the V0C detector. Points corresponding to $q_3$ values are slightly shifted horizontally within each centrality interval for better visibility. Error bars (boxes) indicate statistical (systematic) uncertainties.}
	\label{fig:v42centPbXe_q3}
\end{figure}

In Figs.~\ref{fig:v22centPbXe},~\ref{fig:v32centPbXe}, and~\ref{fig:v42centPbXe}, the charged-particle $v_{2}\{2\}$, $v_{3}\{2\}$, and $v_{4}\{2\}$ are presented for Pb--Pb and Xe--Xe collisions. The flow coefficients are shown for the unbiased sample and the event-shape-selected events. In general, the ESE provides access to the widest range of anisotropic flow values in midcentral and midperipheral events, whereas in central collisions, the smaller average elliptic flow corresponds to a narrower range of selected $v_2$ and $\epsilon_2$ values, reducing the selective power of $q_2$ on the initial eccentricity. For Xe--Xe, it is also worth noting that the $q_2$-selected $v_2\{2\}$ has good statistical precision, while $v_3\{2\}$ and $v_4\{2\}$ have much larger statistical uncertainties.
In the most central Xe--Xe collisions, the $v_3\{2\}$ values show a possible stronger sensitivity to the $q_2$ selection than in the other centrality classes, although the statistical uncertainties remain sizeable. For $v_4\{2\}$, no significant dependence on $q_2$ can be established within the present uncertainties. The behavior in central Xe--Xe collisions may be influenced by the intrinsic deformation of the Xe nucleus, which can affect the initial eccentricity in head-on collisions.
Figures~\ref{fig:v22centPbXe_q3},~\ref{fig:v32centPbXe_q3}, and~\ref{fig:v42centPbXe_q3} present
charged-particle $v_{2}\{2\}$, $v_{3}\{2\}$, and $v_{4}\{2\}$ for Pb--Pb and Xe--Xe collisions using $q_3$–selection. $v_2\{2\}$ for both systems shows a small decrease from low to high $q_3$, but otherwise little sensitivity to the selection. The $v_3\{2\}$ values show strong sensitivity to $q_3$, which is expected as they are driven by the same underlying initial triangularity. Finally, $v_4\{2\}$ shows no significant dependence on $q_3$ beyond the statistical and systematic uncertainties.

\begin{figure}[H]
	\centering
	\begin{minipage}{.49\textwidth} 
		\centering
		\includegraphics[width=\textwidth]{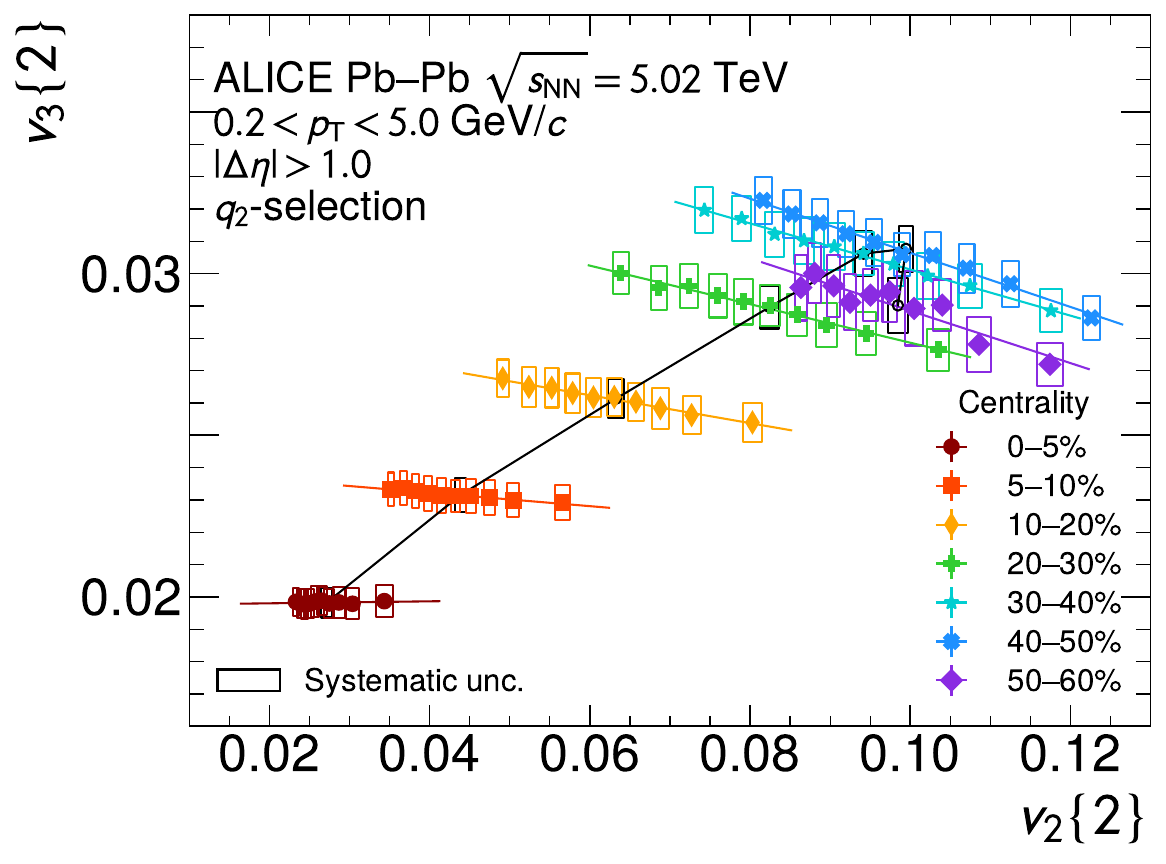}
	\end{minipage}
	\begin{minipage}{.49\textwidth}
		\centering
		\includegraphics[width=\textwidth]{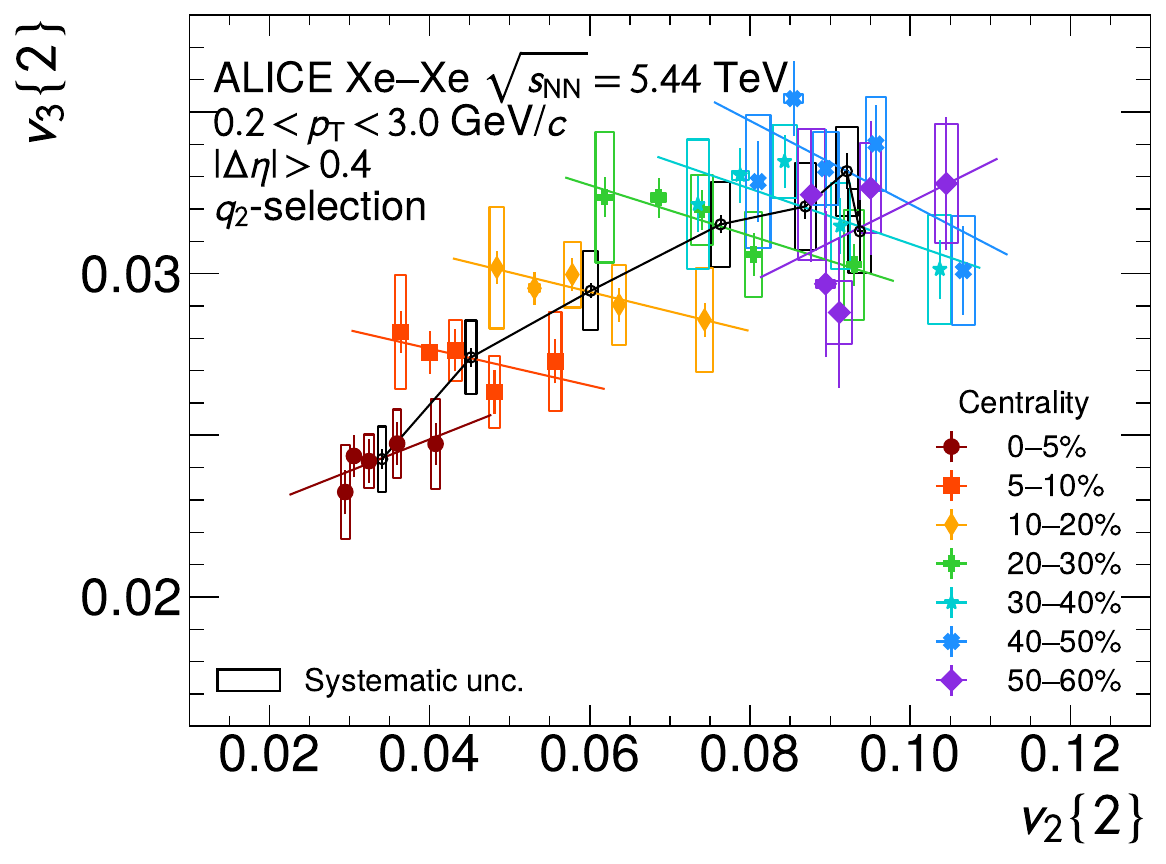}
	\end{minipage}
	\caption{The correlation between $v_3\{2\}$ and $v_2\{2\}$ using $q_2^\mathrm{V0C}$ selection in Pb--Pb (left) and Xe--Xe (right) collisions. The open black points, connected by a black line for visibility, show the $q_2$-unbiased correlation from central to peripheral collisions. The $q_2$-differential correlations, with the lowest (highest) $v_2\{2\}$ value corresponding to 0–-10\% (90–-100\%) $q_2$, are shown by the full markers and linear fits in each centrality interval. Error bars (boxes) indicate statistical (systematic) uncertainties.}
	\label{fig:v2v3cor}
\end{figure}

Figure~\ref{fig:v2v3cor} shows the correlation between $v_2\{2\}$ and $v_3\{2\}$ in centrality and $q_2$ intervals for both Pb--Pb and Xe--Xe collisions. The unbiased $v_2\{2\}$--$v_3\{2\}$ correlation, without ESE selection, changes slope as a function of centrality, indicating a non-trivial relationship between $v_2\{2\}$ and $v_3\{2\}$ that evolves with centrality. This trend has also been previously observed by the ATLAS Collaboration~\cite{ATLAS:2015qwl} in Pb--Pb collisions. When the events are classified into $q_2$ intervals within each centrality range, a clear linear anticorrelation between $v_2\{2\}$ and $v_3\{2\}$ becomes apparent. This anticorrelation is particularly strong in peripheral collisions, in agreement with prior measurements~\cite{ATLAS:2015qwl,ALICE:2016kpq, ALICE:2021adw}. The anticorrelation can also be understood from a physical standpoint: a stronger elliptic shape of the system geometry suppresses the magnitude of a possible triangular anisotropy. In the most central collisions, there is no visible anticorrelation observed, likely due to the more isotropic shape of the initial geometry.

\begin{figure}[H]
		\centering
        \centering
    	\begin{minipage}{.49\textwidth}
    		\centering
    		\includegraphics[width=\textwidth]{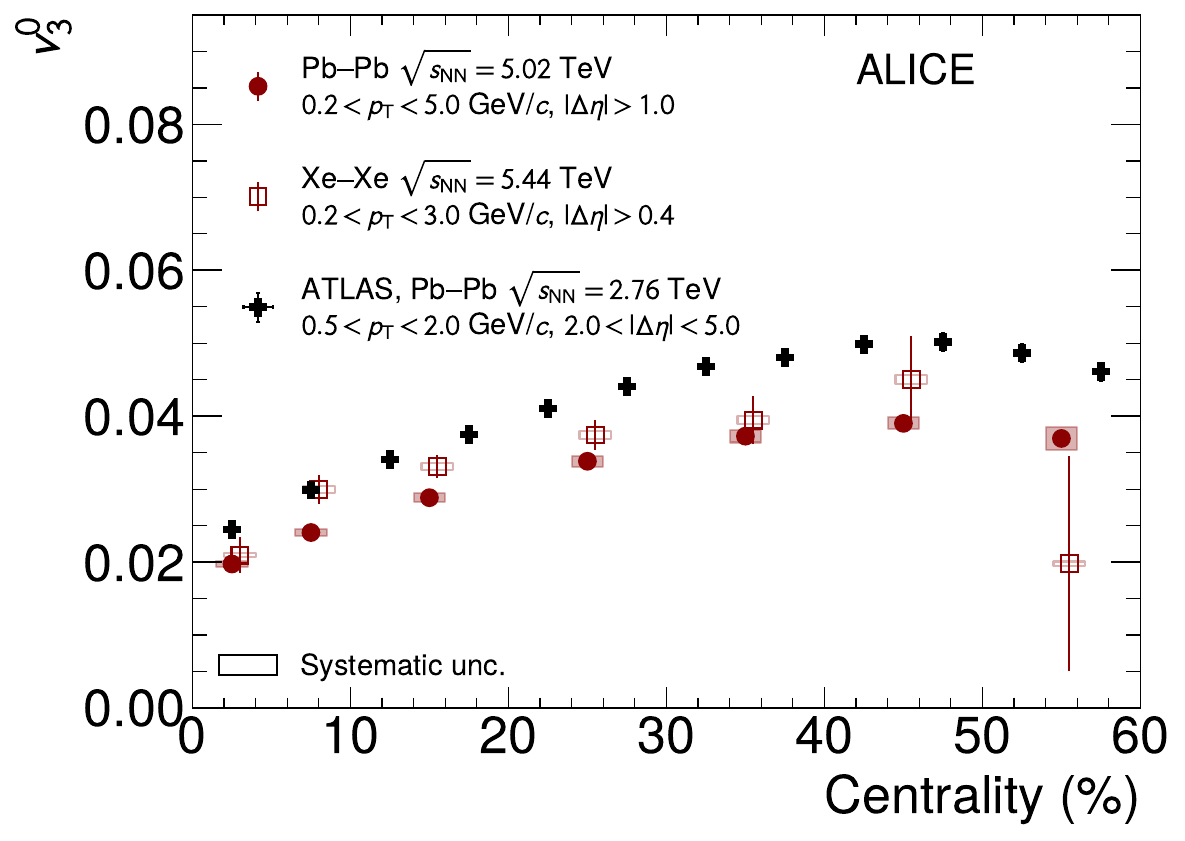}
    	\end{minipage}
    	\begin{minipage}{.49\textwidth}
    		\centering
    		\includegraphics[width=\textwidth]{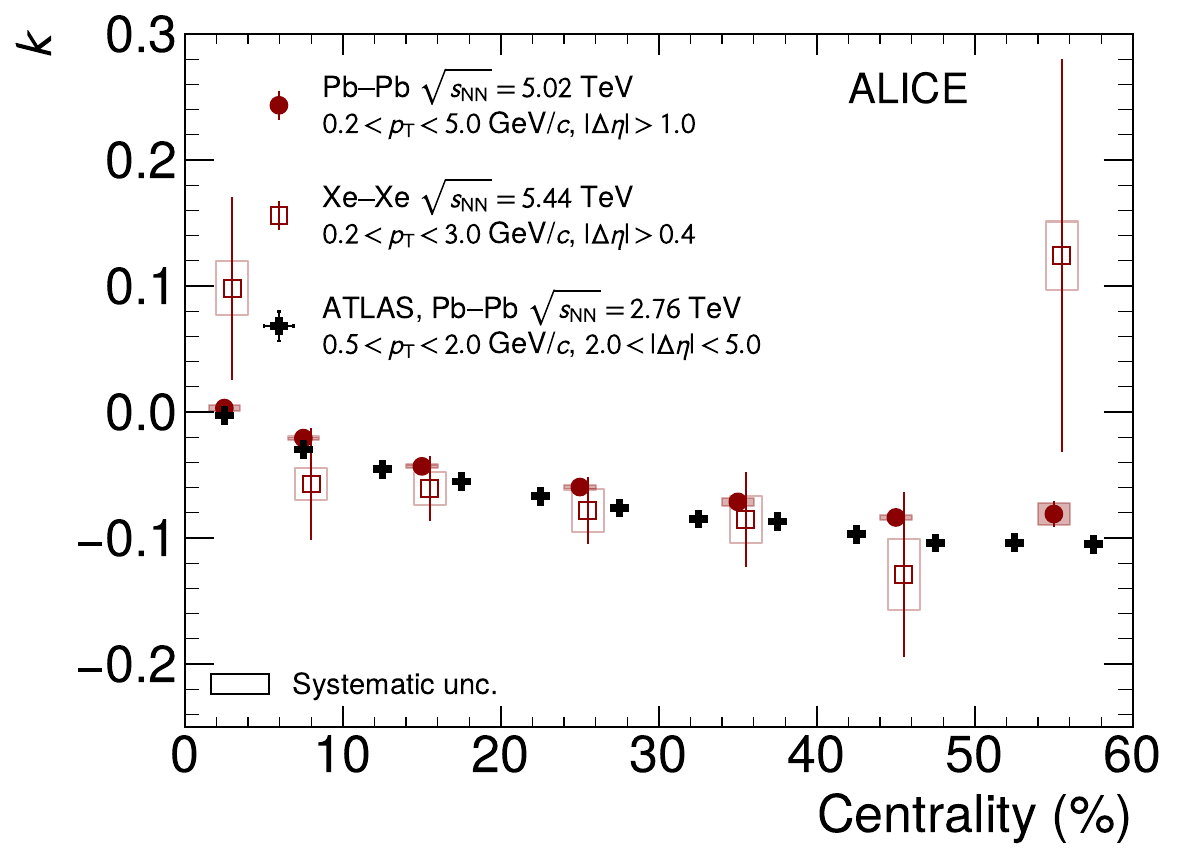}
    	\end{minipage}
		\caption{The coefficients $v_3^0$ (left) and $k$ (right) from fits of Eq.~\eqref{eq:v3v2} to the $v_2\{2\}$--$v_3\{2\}$ correlation shown in Fig.~\ref{fig:v2v3cor} for Pb--Pb (solid red markers) and Xe--Xe (open red markers) collisions as a function of the collision centrality. Xe--Xe points are slightly shifted horizontally. Measurements from the ATLAS Collaboration in Ref.~\cite{ATLAS:2015qwl} (black points) are shown for charged particles in the kinematic range $0.5 < p_\mathrm{T} < 2.0~\text{GeV}/c$ and $|\eta| < 2.5$. Error bars (boxes) indicate statistical (systematic) uncertainties.}
		\label{fig:v2v3cor_fit}
\end{figure}

The data in Fig.~\ref{fig:v2v3cor} are fitted with a linear function of $v_3\{2\}$ on $v_2\{2\}$ (Eq.~\ref{eq:v3v2}). These fits show clear negative slopes in most centrality ranges, reflecting the strong anticorrelation between $v_2\{2\}$ and $v_3\{2\}$, which is consistent with the negative SC(3,2) values measured in previous symmetric cumulants studies~\cite{ALICE:2016kpq,ALICE:2021adw,ALICE:2024nqd}. The fitted intercept $v_3^0$ and slope $k$ obtained for Pb--Pb and Xe--Xe collisions are shown in Fig.~\ref{fig:v2v3cor_fit} and are compared to measurements from the ATLAS Collaboration. The ALICE values differ slightly from those measured by the ATLAS Collaboration~\cite{ATLAS:2015qwl}, likely due to the different kinematic selections of the analysed particles, i.e., the minimum-$p_\mathrm{T}$ selection criterion. The magnitude and behavior of the measured $v_2\{2\}$--$v_3\{2\}$ correlations are similar in Pb--Pb and Xe--Xe, except in the most central collisions. The differences in the central region may originate from the different nuclear shapes of Pb and Xe, which define the initial geometry in head-on collisions. As a result, the $v_2\{2\}$--$v_3\{2\}$ correlation is stronger in Xe--Xe than in Pb--Pb collisions at similar centrality, and even changes sign, which is also seen in models at very high multiplicity~\cite{PhysRevLett.123.142301}. In peripheral Xe--Xe collisions, an anticorrelation between $v_2\{2\}$ and $v_3\{2\}$ cannot be established because of the large statistical uncertainties on $v_3\{2\}$.

\begin{figure}[H]
	\centering
	\begin{minipage}{.49\textwidth}
		\centering
		\includegraphics[width=\textwidth]{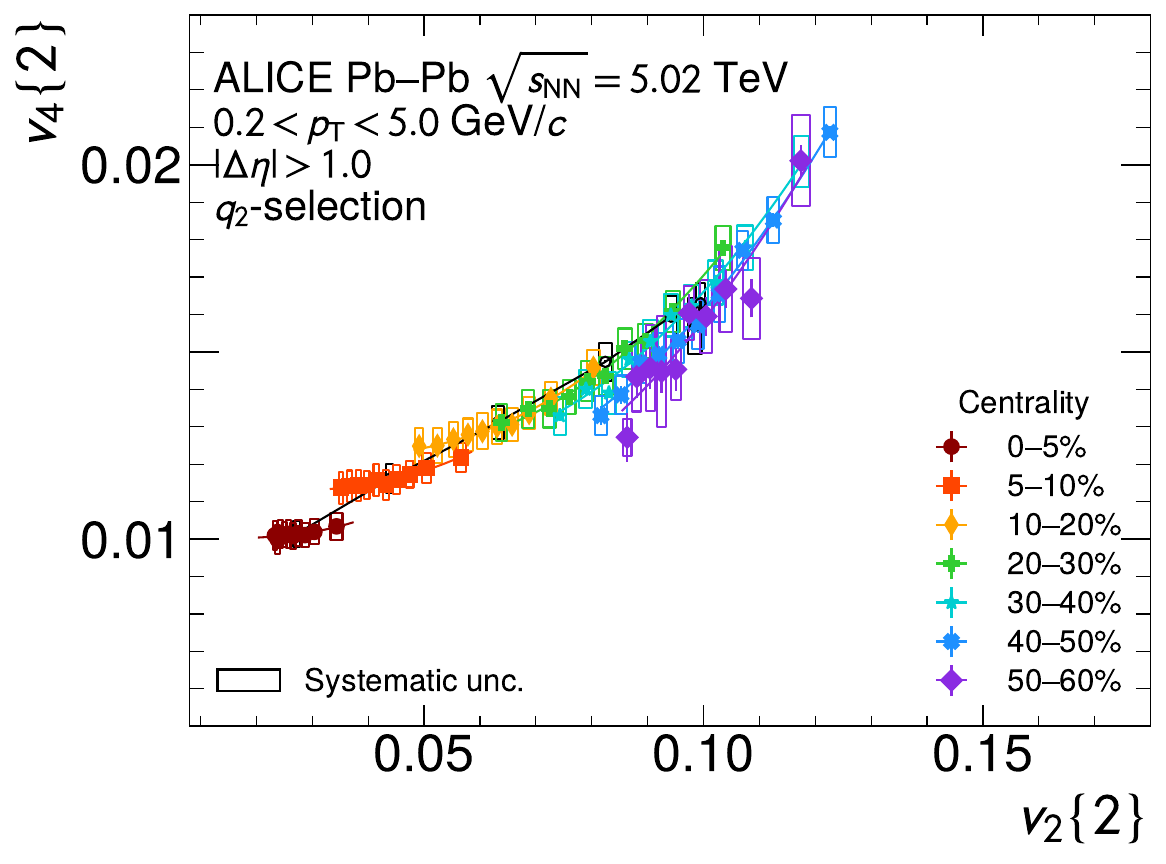}
	\end{minipage}
	\begin{minipage}{.49\textwidth}
		\centering
		\includegraphics[width=\textwidth]{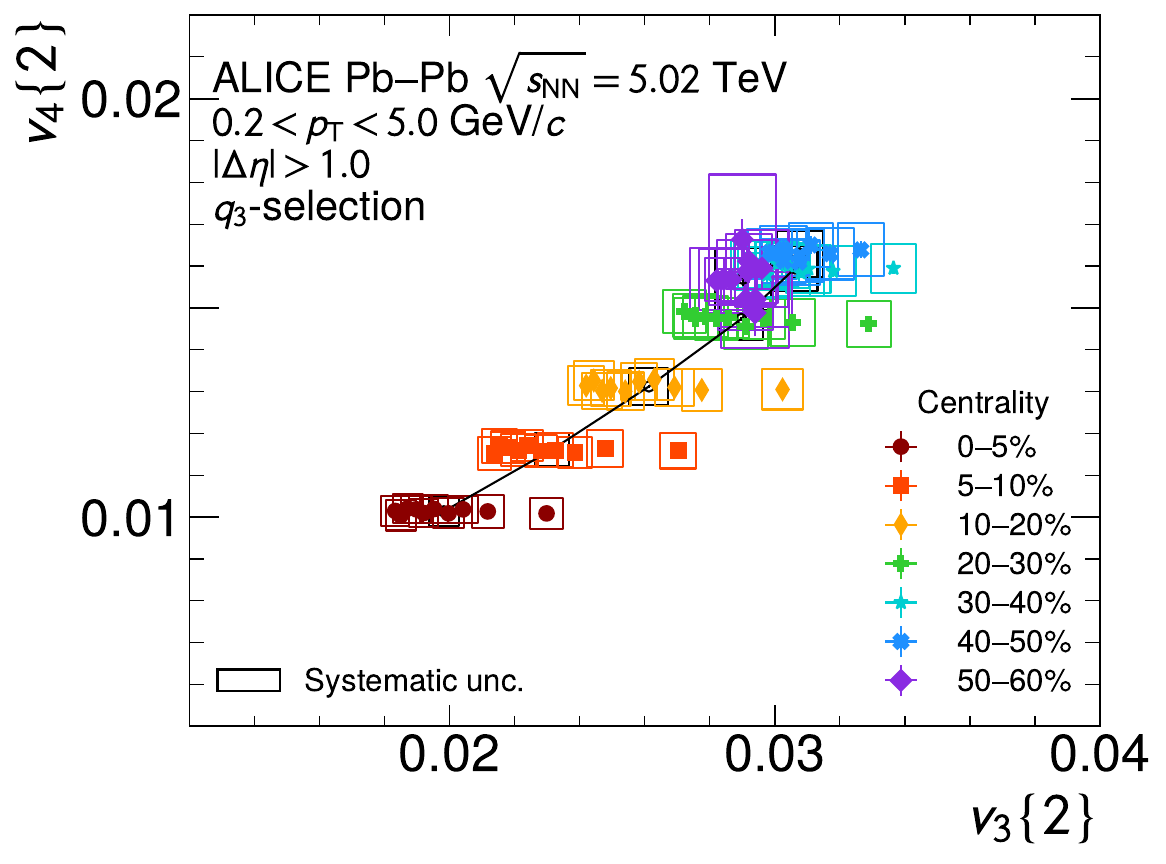}
	\end{minipage}
	\caption{The correlation between $v_4\{2\}$ and $v_2\{2\}$ using $q_2^\mathrm{V0C}$ selection (left) and $v_4\{2\}$ and $v_3\{2\}$ using $q_3^\mathrm{V0C}$ selection (right) in Pb--Pb collisions. The open black points, connected by a black line for visibility, show the unbiased correlation. The $q_n$-differential correlations are shown by the full markers in each centrality interval. Error bars (boxes) indicate statistical (systematic) uncertainties.}
	\label{fig:v2v4cor}
\end{figure}

Figure~\ref{fig:v2v4cor} (left) shows the correlation between $v_2\{2\}$ and $v_4\{2\}$ for Pb--Pb collisions using $q_2$ selection. These correlations provide insight into how the collision geometry influences higher-order flow harmonics. The unbiased $v_2\{2\}$--$v_4\{2\}$ correlation changes slope as a function of centrality, indicating a strong centrality dependence and a complex interplay of $v_2\{2\}$ and $v_4\{2\}$, in particular in semi-peripheral collisions, where elliptic flow becomes dominant. Within each centrality interval, however, $v_4\{2\}$ increases monotonically as a function of $v_2\{2\}$.  The observed positive correlation is also consistent with the measurements of symmetric cumulants SC(4,2)~\cite{ALICE:2016kpq, ALICE:2021adw, ALICE:2017kwu}. In peripheral collisions, the dependence of $v_4\{2\}$ on $v_2\{2\}$ becomes increasingly nonlinear. In both the $v_2\{2\}$--$v_3\{2\}$ and $v_2\{2\}$--$v_4\{2\}$ correlation studies, symmetric cumulants provide valuable validation of the trends observed using ESE.

The positive nonlinear correlation observed between $v_2\{2\}$ and $v_4\{2\}$ in Pb--Pb collisions is largely driven by the nonlinear part of $v_4\{2\}$ which depends on the elliptic eccentricity ($\epsilon_2$). This effect is expected from hydrodynamic models, where $v_4\{2\}$ is known to be partially driven by the second-order eccentricity in the initial-state geometry~\cite{Yan:2015jma}, and has been confirmed by previous measurements~\cite{ALICE:2017fcd, ALICE:2020sup}. 

In Fig.~\ref{fig:v2v4cor} (right), the correlation between $v_3\{2\}$ and $v_4\{2\}$ is presented using ESE with $q_3$ selection. Within each centrality interval, the data exhibit a tendency towards an anticorrelation, consistent with previous SC(4,3) measurements~\cite{ALICE:2021adw}, although the effect is not statistically significant. Given that $v_4^2$ is positively correlated with $v_2^2$, and $v_2^2$ is anti-correlated with $v_3^2$, it is reasonable to expect that $v_3\{2\}$ and $v_4\{2\}$ would exhibit an anticorrelation. This reflects the intricate relationship between different harmonic orders, where the influence of $v_2\{2\}$ on $v_4\{2\}$ contributes to the overall structure of these correlations in heavy-ion collisions.

\begin{figure}[H]
        \centering
    	\begin{minipage}{.49\textwidth}
    		\centering
    		\includegraphics[width=\textwidth]{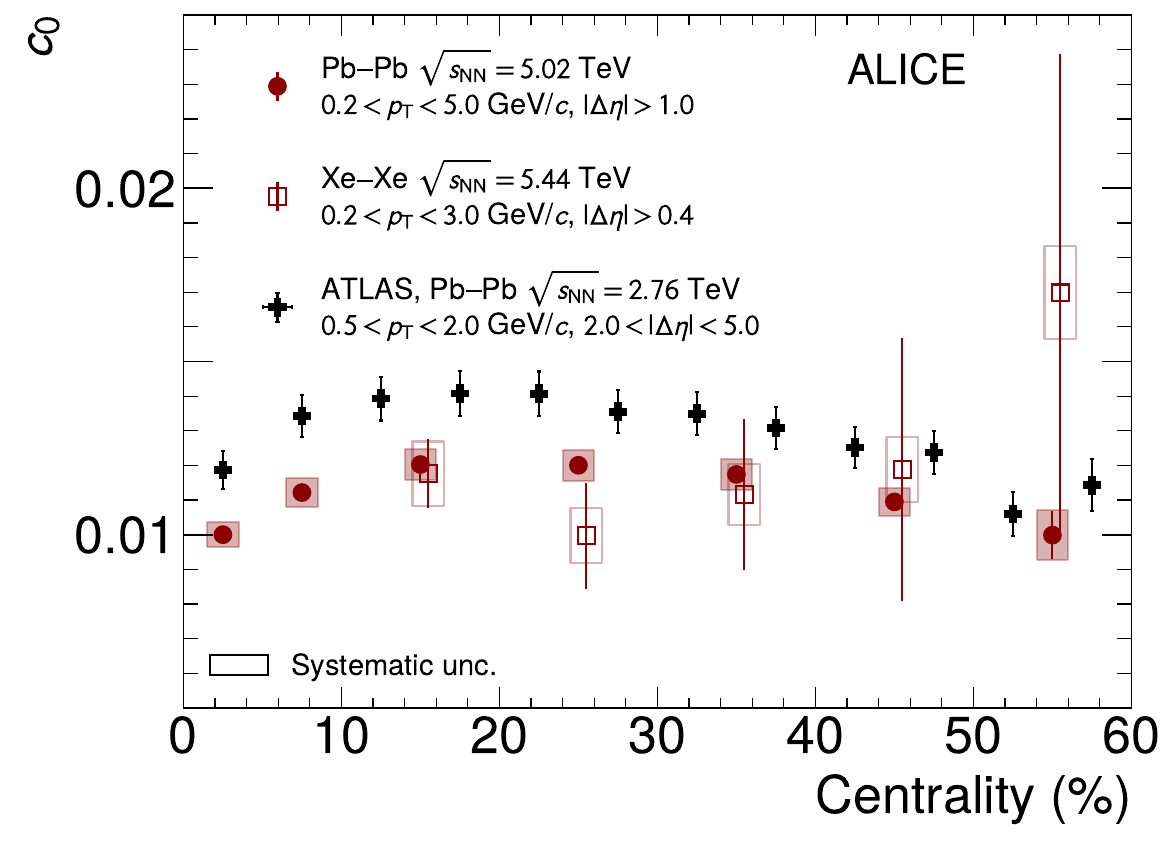}
    	\end{minipage}
    	\begin{minipage}{.46\textwidth}
    		\centering
    		\includegraphics[width=\textwidth]{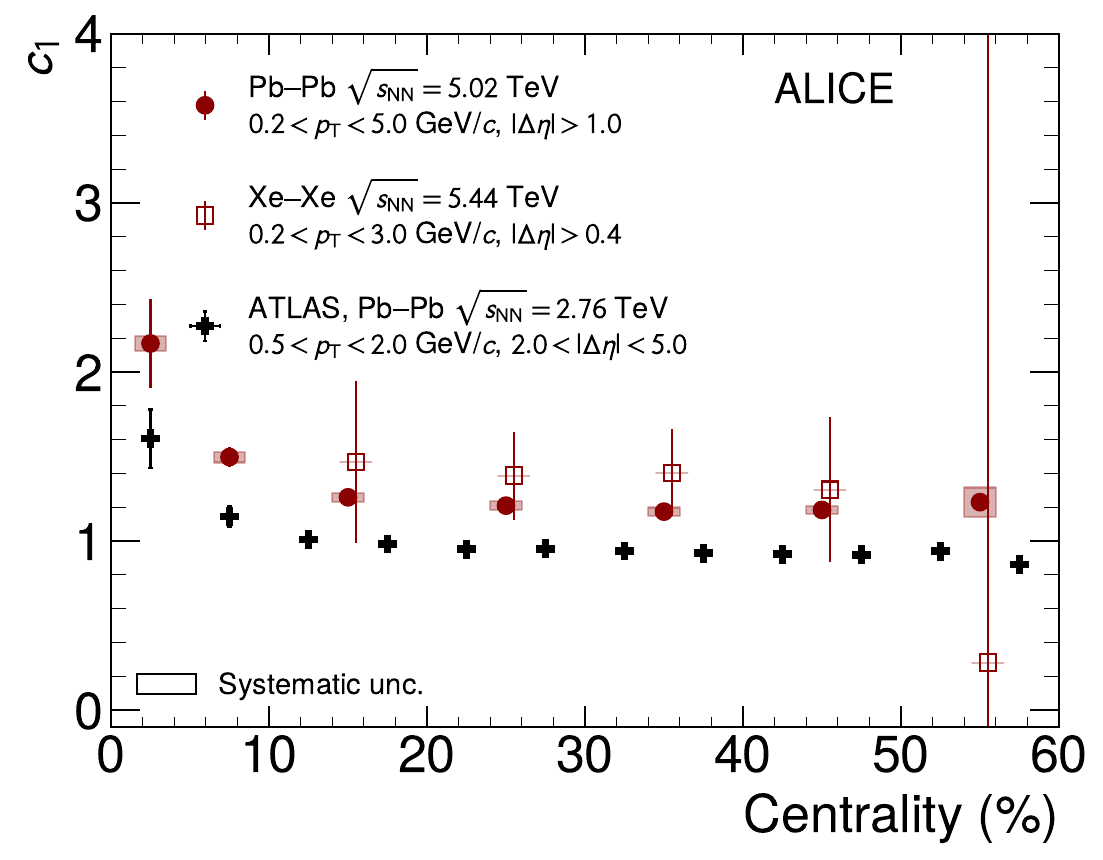}
    	\end{minipage}
		\caption{The coefficients $c_0$ (left), representing the linear contribution, and $c_1$ (right), quantifying the nonlinear dependence on $v_2^2$, from fits of Eq.~\eqref{eq:v4fit} to the $v_2\{2\}$--$v_4\{2\}$ correlations shown in Fig.~\ref{fig:v2v4cor} for Pb--Pb (solid red markers) and Xe--Xe (open red markers) collisions as a function of the collision centrality. Xe--Xe points are slightly shifted horizontally for better visibility. Measurements from the ATLAS Collaboration in Ref.~\cite{ATLAS:2015qwl} (black points) are shown for charged particles in the kinematic range $0.5 < p_\mathrm{T} < 2.0~\text{GeV}/c$ and $|\eta| < 2.5$. Error bars (boxes) indicate statistical (systematic) uncertainties.}
		\label{fig:v2v4cor_fit}
\end{figure}

To quantify the relationship between $v_2\{2\}$ and $v_4\{2\}$, the data in Fig.~\ref{fig:v2v4cor} (left) is fit with a functional form that separates a linear contribution, $c_0$, from a nonlinear contribution given by $c_1 v_2^2$ (Eq.~\ref{eq:v4fit}). The resulting fit coefficients in Pb--Pb collisions and Xe--Xe collisions are presented in Fig.~\ref{fig:v2v4cor_fit}, and compared with measurements from the ATLAS Collaboration in Ref.~\cite{ATLAS:2015qwl}. While the overall trend is consistent with the measurement from the ATLAS Collaboration, a difference in magnitude is observed, possibly due to the different lower-$\pt$ limits. The data reveal a strong nonlinear component, showing that the development of $v_4$ is closely tied to the elliptic flow $v_2$. For Xe--Xe collisions, the statistical uncertainties on $v_4\{2\}$ begin to dominate the correlations, yet they remain consistent with the findings in Pb--Pb.

\begin{figure}[h!]
	\centering
	\includegraphics[width=0.55\textwidth]{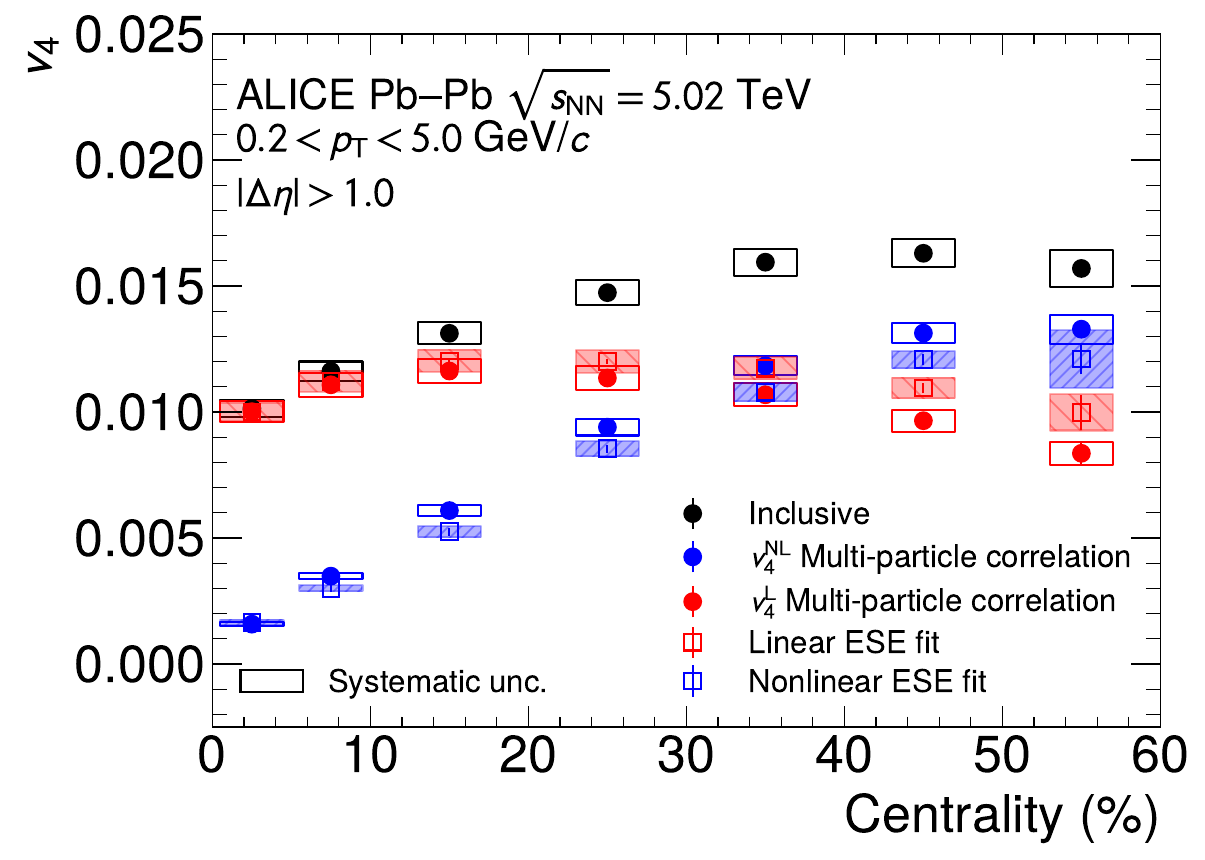}
	\caption{The contribution of the linear and nonlinear flow modes of $v_4$ extracted via multiparticle correlations (solid red and blue points) and using the ESE (open red and blue points), as well as the unbiased $v_4\{2,|\Delta\eta|>1.0\}$ (black points),  as a function of centrality in Pb--Pb collisions. Error bars (boxes) indicate statistical (systematic) uncertainties.}
	\label{fig:v4_lin_nlin}
\end{figure}

The resulting linear and nonlinear parts of $v_4$ are obtained from the fit coefficients by taking $v_4^\mathrm{L}=c_0$ and calculating $v_4^\mathrm{NL}$ from the remaining contribution to the unbiased $v_4^2$ (Eq.~\ref{eq:extractv4nl}), and are shown in Fig.~\ref{fig:v4_lin_nlin}. They are compared with the flow modes obtained from multiparticle correlations (Eq.~\ref{eq:v4nl}), where the nonlinear mode is measured with mixed-harmonic correlations and the linear mode is obtained from the remaining contribution to the total $v_4^2$ (Eqs.~\ref{eq:v4parts} and~\ref{eq:v4nl}). The results from the two approaches agree with the values in Refs.~\cite{ALICE:2017fcd,ALICE:2020sup}. The small differences between the measurements of $v_4^\mathrm{L}$ and $v_4^\mathrm{NL}$ obtained with the two methods may arise from the different correlation orders used in the analysis. The ESE-based measurements rely on two-particle correlations, whereas the alternative method employs multiparticle correlations, resulting in different sensitivities to the underlying flow distribution (see Eqs.~\ref{eq:vn2} and~\ref{eq:vn4}).

\begin{figure}[H]
	\centering
	\includegraphics[width=\textwidth]{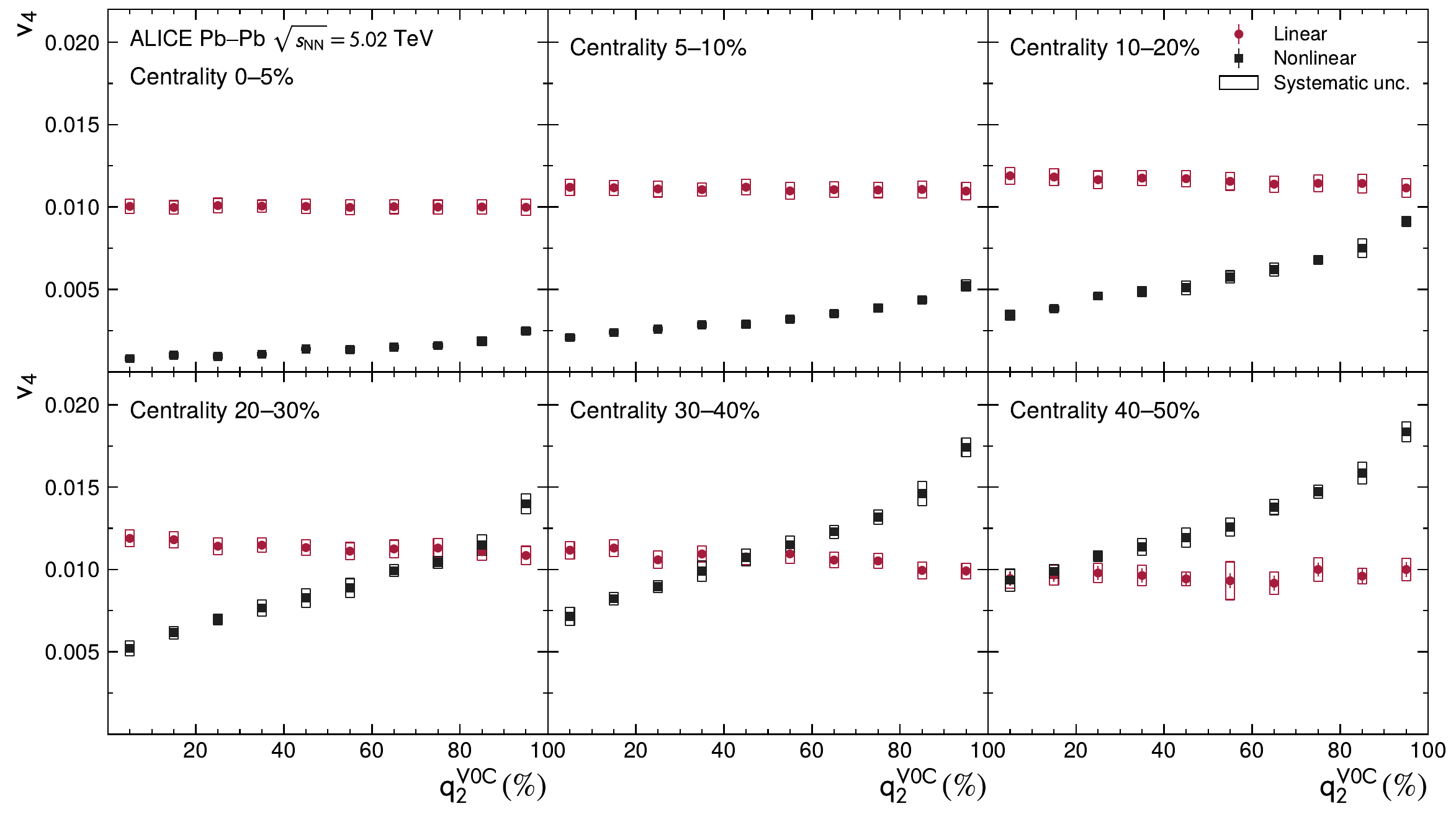}
	\caption{The linear (red) and nonlinear (black) flow modes of $v_4$ measured with multiparticle correlations as a function of the $q_2^\mathrm{V0C}$ percentile in different centrality intervals in Pb--Pb collisions. Error bars (boxes) indicate statistical (systematic) uncertainties.}
	\label{fig:v4coupling}
\end{figure}

To test whether the linear and nonlinear components of $v_4$ have the expected dependence on $\epsilon_2$, $v_4^{\rm L}$ and $v_4^{\rm NL}$ are directly measured using multiparticle cumulants in $q_2$ percentile intervals for the first time. The nonlinear component is obtained from mixed-harmonic correlations, and the linear component from the remaining contribution to $v_4^2$ (Eqs.~\ref{eq:v4parts} and~\ref{eq:v4nl}); the results are shown in Fig.~\ref{fig:v4coupling}. The results clearly demonstrate the separation of the linear and nonlinear parts of $v_4$, with $v_4^{\rm L}$ measured through multi-particle cumulants displaying no dependence on $q_2$, in accordance with the expectations, since it is primarily determined by the fourth order initial eccentricity, $\epsilon_4$.  Measuring the linear and nonlinear components as a function of $q_2$ allows us to put constraints on the extent to which the fourth-order initial cumulant-defined eccentricity coefficient, $\epsilon_4'$, is influenced by the elliptic flow eccentricity $\epsilon_2$.  $v_4^{\rm NL}$ displays a strong $q_2$-dependence, except for in the most central collisions in which the geometrical anisotropy is weaker. The observed $q_2$ dependence is consistent with the expectations that the nonlinear flow modes are driven by the squared initial-state ellipticity, $\epsilon_2^2$. These measurements provide an independent test of this coupling and support the conclusions in Ref.~\cite{ALICE:2017fcd}.

\begin{figure}[H]
	\centering
	\includegraphics[width=\textwidth]{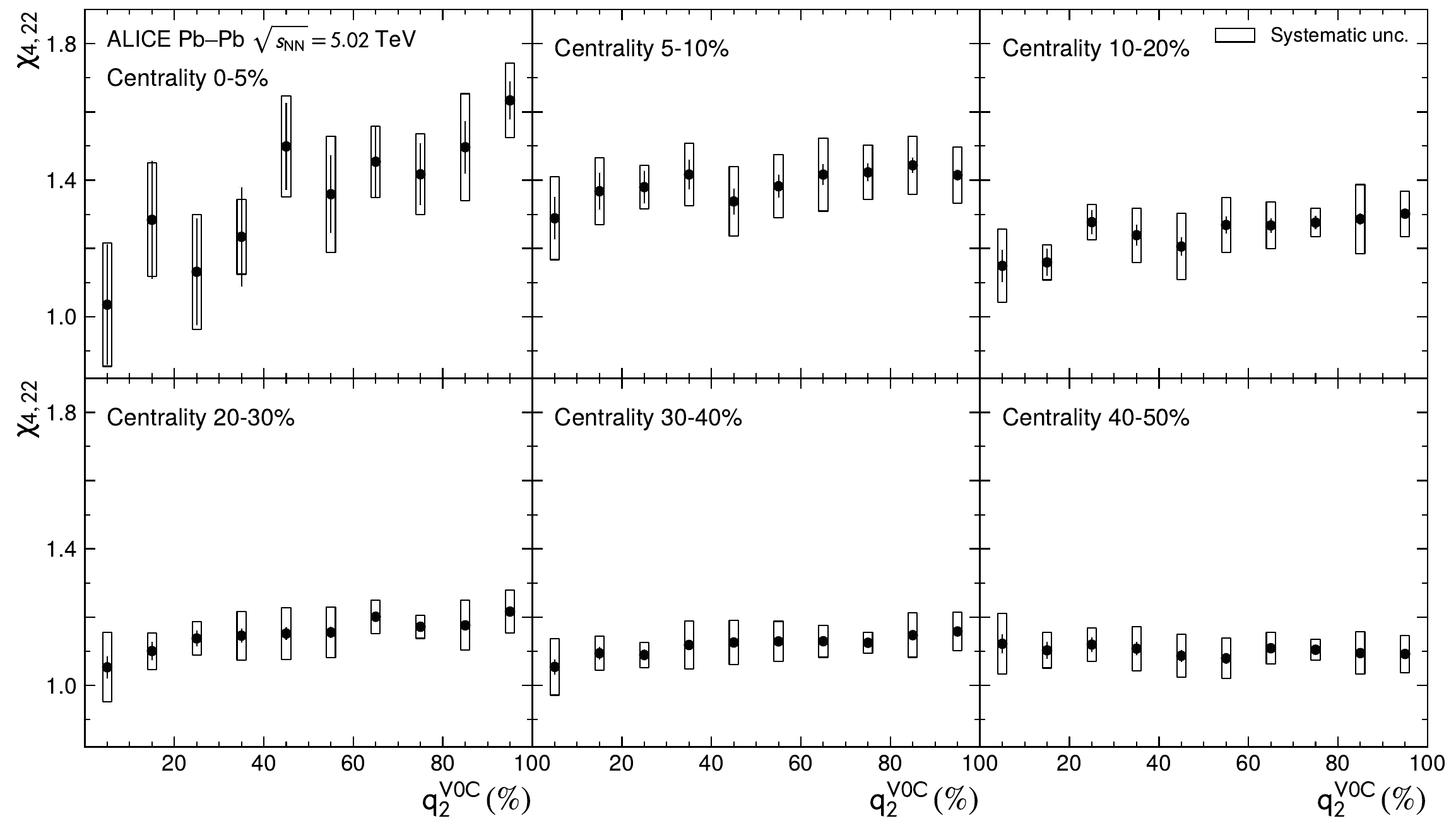}
	\caption{The nonlinear response coefficient $\chi_{4,22}$ as a function of $q_2^\mathrm{V0C}$ percentile in different centrality intervals in Pb--Pb collisions. Error bars (boxes) indicate statistical (systematic) uncertainties.}
	\label{fig:chi422}
\end{figure}

In Fig.~\ref{fig:chi422}, the coupling is further investigated by evaluating the nonlinear response coefficient $\chi_{4,22}$ with the multiparticle cumulant method, where the nonlinear $v_4$ mode is normalized by $v_2^2$ (Eqs.~\ref{eq:v4nl} and~\ref{eq:chi422}). $\chi_{4,22}$ shows a flat dependence on $q_2$ in all centrality intervals except 0--5\% centrality, where an increase with $q_2$ is observed. Therefore, it largely confirms that $v_4^{\rm NL}\propto v_2^2$ and $\chi_{4,22}$ does not depend on the initial geometry $\varepsilon_2$, which is also one of the model assumptions in Ref.~\cite{Yan:2015jma}.

\begin{figure}[H]
	\centering
	\begin{minipage}{.49\textwidth}
		\centering
	\includegraphics[width=\textwidth]{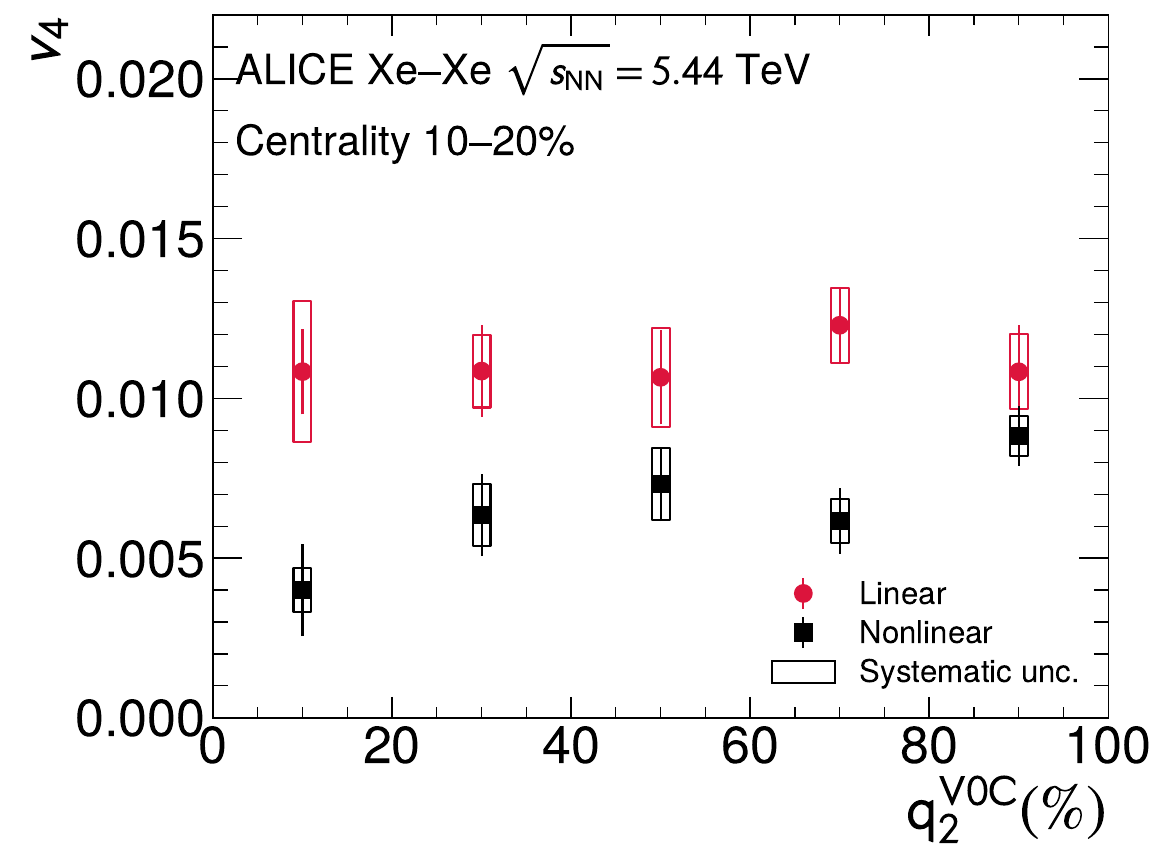}
	\end{minipage}
	\begin{minipage}{.49\textwidth}
		\centering
	\includegraphics[width=\textwidth]{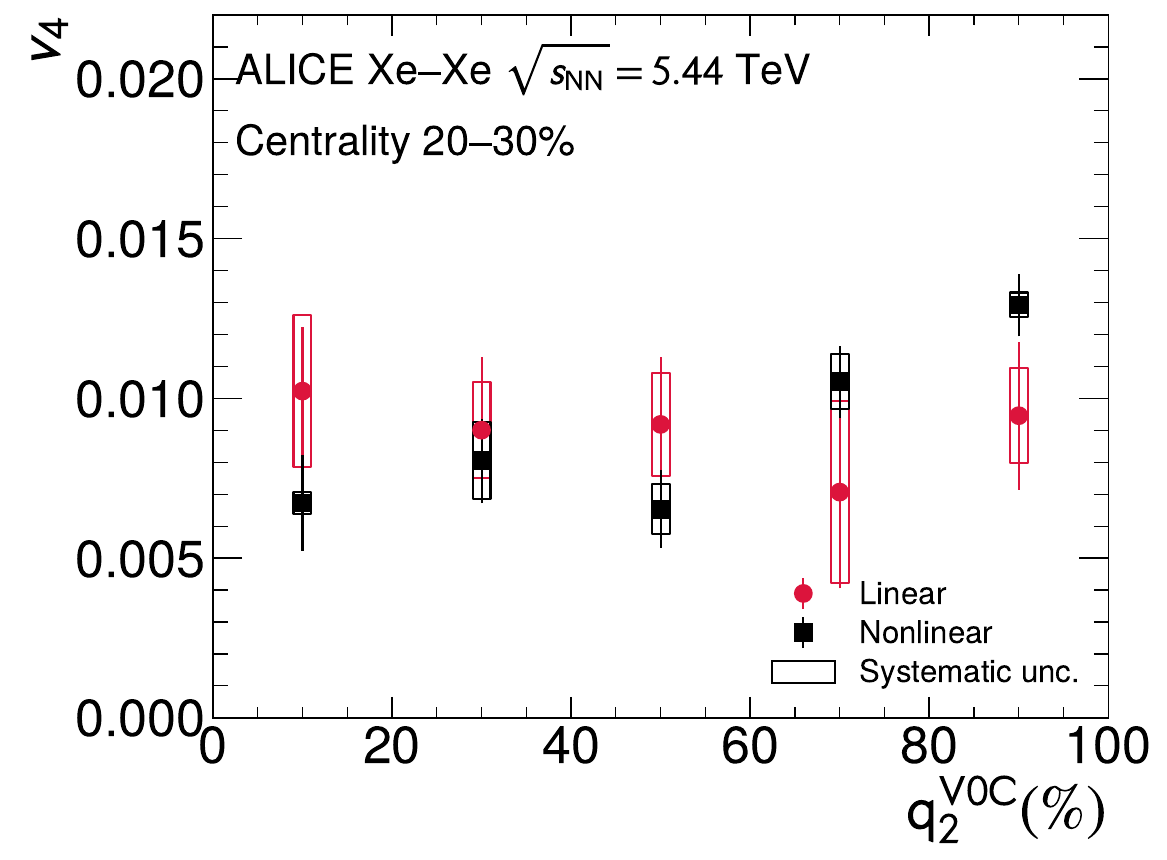}
	\end{minipage}
	\caption{The linear (red) and nonlinear (black) flow modes of $v_4$ measured with multiparticle correlations as a function of the $q_2^\mathrm{V0C}$ percentile in 10--20\% (left) and 20--30\% (right) central Xe--Xe collisions. Error bars (boxes) indicate statistical (systematic) uncertainties.}
	\label{fig:v4couplingXe_2}
\end{figure}

The linear and nonlinear components of $v_4$, $v_4^\mathrm{L}$ and $v_4^{\rm NL}$, are measured for the first time using multiparticle correlations in Xe--Xe collisions at $\sqrt{s_{\rm NN}} = 5.44$ TeV and are shown as a function of $q_2$ in Fig.~\ref{fig:v4couplingXe_2}. The behavior observed in Xe--Xe collisions is qualitatively similar to that seen in Pb--Pb collisions: The linear part exhibits a weak dependence on $q_2$ across the centrality range of 10--30\%, while the nonlinear part shows a stronger increase with $q_2$. This result reaffirms the role of $\epsilon_2$ in driving the nonlinear contribution to $v_4$. The monotonic increase in $v_4^{\rm NL}$ with $q_2$ in both Pb--Pb and Xe--Xe collisions further supports the conclusion that the nonlinear flow component is sensitive to initial elliptic fluctuations, while the linear component remains largely independent of these fluctuations.

\begin{figure}[H]
	\centering
	\includegraphics[width=0.6\textwidth]{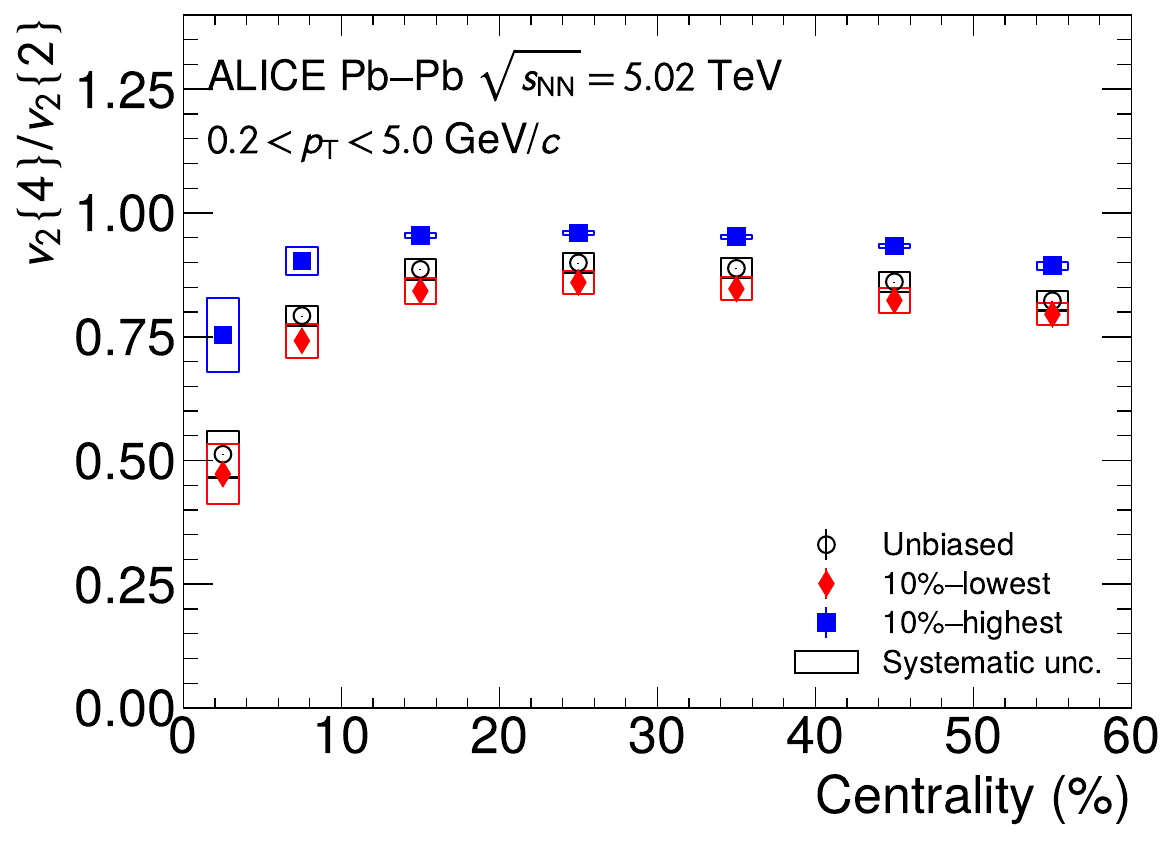}
	\caption{The ratio of $v_2\{4\}$ and $v_2\{2\}$ in the 10\% most elliptic events (blue), 10\% least elliptic events (red), and unbiased events (black), as a function of centrality in Pb--Pb collisions. Error bars (boxes) indicate statistical (systematic) uncertainties.}
	\label{fig:v2fluct}
\end{figure}

In addition to studying correlations between various flow harmonics presented above, the ESE technique can also be used with multiparticle correlations to probe event-by-event fluctuations of the collective flow. 
Figure~\ref{fig:v2fluct} shows the ratio of $v_2\{4\}$ to $v_2\{2\}$ as a function of centrality for the 10\% lowest and 10\% highest $q_2$ percentiles, as well as for the unbiased sample. While the overall trend with centrality remains consistent between $q_2$ selections, the magnitude of the ratio varies.

The results indicate that events with the highest $q_2$ exhibit smaller flow fluctuations than the unbiased sample. This is consistent with a more clearly defined initial-state geometry in events with large eccentricity. When considering the lowest-$q_2$ events, the geometry should also be rather well defined, but these points seem to exhibit slightly larger flow fluctuations than the unbiased data points. Thus, while the ESE selection is sensitive to eccentricity, it can also modify the flow fluctuations and may introduce additional selection biases. The change in the event-by-event flow fluctuations due to the $q_2$ selection might be the cause of the increase of $\chi_{4,22}$ as a function of $q_2$ in the $0$--$5\%$ centrality range in Fig.~\ref{fig:chi422}.

\section{Conclusion}
In summary, this paper presents a comprehensive study of the correlations between the flow harmonics, $v_2$, $v_3$, and $v_4$, using Event Shape Engineering selections on $q_2$ and $q_3$ in Pb--Pb and Xe--Xe collisions at the LHC. The results qualitatively agree with the normalized symmetric cumulants SC(3,2) and SC(4,2), demonstrating a more direct way to understand the correlations by investigating sensitivity to the initial geometry of collisions. Using a combination of multiparticle correlations and ESE with selections based on the flow vector $q_2$, the linear and nonlinear components of $v_4$ are separated. While the nonlinear component exhibits a clear dependence on the $q_2$ selection, the linear component remains largely unchanged, consistent with the two components being uncorrelated. Additionally, the assumption that $\chi_{4,22}$ remains independent of $\varepsilon_2$ is largely confirmed, although some dependence on $q_2$ is observed in the most central ($0$--$5\%$) collisions. The results, compared across experiments and collision systems, also reveal a slightly stronger sensitivity to $q_2$ in Xe--Xe collisions, consistent with the hypothesis that deformed nuclei have a better-defined geometry in more central collisions, which may allow more precise event selections based on the initial geometry in systems with deformed nuclei. Finally, ESE is also used to study the event-by-event fluctuations of $v_2$, revealing that not only does ESE impose a selection on geometry, but also impacts the magnitude of the flow fluctuations. 

Overall, the findings deepen understanding of the role of initial-state geometric anisotropies in the development of anisotropic flow harmonics in heavy-ion collisions. The consistency of the findings across both Pb--Pb and Xe--Xe systems reinforces the robustness of ESE and flow harmonic correlations as tools for probing the properties of the QGP and its sensitivity to the initial geometry of the collisions. More importantly, these results demonstrate the unique capability of ESE to experimentally control the initial geometry of heavy-ion collisions, providing a powerful approach to disentangle the effects of the initial conditions from the subsequent collective evolution of the QGP. This makes ESE a valuable tool for future precision studies of both the initial-state geometry and the medium response in relativistic nuclear collisions at the LHC.

\newenvironment{acknowledgement}{\relax}{\relax}
\begin{acknowledgement}
\section*{Acknowledgements}
% Version: 2026-10-01

The ALICE Collaboration would like to thank all its engineers and technicians for their invaluable contributions to the construction of the experiment and the CERN accelerator teams for the outstanding performance of the LHC complex.
The ALICE Collaboration gratefully acknowledges the resources and support provided by all Grid centres and the Worldwide LHC Computing Grid (WLCG) collaboration.
The ALICE Collaboration acknowledges the following funding agencies for their support in building and running the ALICE detector:
A. I. Alikhanyan National Science Laboratory (Yerevan Physics Institute) Foundation (ANSL), State Committee of Science and World Federation of Scientists (WFS), Armenia;
Austrian Academy of Sciences, Austrian Science Fund (FWF): [M 2467-N36] and Nationalstiftung f\"{u}r Forschung, Technologie und Entwicklung, Austria;
Ministry of Communications and High Technologies, National Nuclear Research Center, Azerbaijan;
Rede Nacional de Física de Altas Energias (Renafae), Financiadora de Estudos e Projetos (Finep), Funda\c{c}\~{a}o de Amparo \`{a} Pesquisa do Estado de S\~{a}o Paulo (FAPESP) and The Sao Paulo Research Foundation  (FAPESP), Brazil;
Bulgarian Ministry of Education and Science, within the National Roadmap for Research Infrastructures 2020-2027 (object CERN), Bulgaria;
Ministry of Education of China (MOEC) , Ministry of Science \& Technology of China (MSTC) and National Natural Science Foundation of China (NSFC), China;
Ministry of Science and Education and Croatian Science Foundation, Croatia;
Centro de Aplicaciones Tecnol\'{o}gicas y Desarrollo Nuclear (CEADEN), Cubaenerg\'{\i}a, Cuba;
Ministry of Education, Youth and Sports of the Czech Republic, Czech Republic;
The Danish Council for Independent Research | Natural Sciences, the VILLUM FONDEN and Danish National Research Foundation (DNRF), Denmark;
Helsinki Institute of Physics (HIP), Finland;
Commissariat \`{a} l'Energie Atomique (CEA) and Institut National de Physique Nucl\'{e}aire et de Physique des Particules (IN2P3) and Centre National de la Recherche Scientifique (CNRS), France;
Bundesministerium f\"{u}r Forschung, Technologie und Raumfahrt (BMFTR) and GSI Helmholtzzentrum f\"{u}r Schwerionenforschung GmbH, Germany;
National Research, Development and Innovation Office, Hungary;
Department of Atomic Energy Government of India (DAE), Department of Science and Technology, Government of India (DST), University Grants Commission, Government of India (UGC) and Council of Scientific and Industrial Research (CSIR), India;
National Research and Innovation Agency - BRIN, Indonesia;
Istituto Nazionale di Fisica Nucleare (INFN), Italy;
Japanese Ministry of Education, Culture, Sports, Science and Technology (MEXT) and Japan Society for the Promotion of Science (JSPS) KAKENHI, Japan;
Consejo Nacional de Ciencia (CONACYT) y Tecnolog\'{i}a, through Fondo de Cooperaci\'{o}n Internacional en Ciencia y Tecnolog\'{i}a (FONCICYT) and Direcci\'{o}n General de Asuntos del Personal Academico (DGAPA), Mexico;
Nederlandse Organisatie voor Wetenschappelijk Onderzoek (NWO), Netherlands;
The Research Council of Norway, Norway;
Pontificia Universidad Cat\'{o}lica del Per\'{u}, Peru;
Ministry of Science and Higher Education, National Science Centre and WUT ID-UB, Poland;
Korea Institute of Science and Technology Information and National Research Foundation of Korea (NRF), Republic of Korea;
Ministry of Education and Scientific Research, Institute of Atomic Physics, Ministry of Research and Innovation and Institute of Atomic Physics and Universitatea Nationala de Stiinta si Tehnologie Politehnica Bucuresti, Romania;
Ministerstvo skolstva, vyskumu, vyvoja a mladeze SR, Slovakia;
National Research Foundation of South Africa, South Africa;
Swedish Research Council (VR) and Knut \& Alice Wallenberg Foundation (KAW), Sweden;
European Organization for Nuclear Research, Switzerland;
Suranaree University of Technology (SUT), National Science and Technology Development Agency (NSTDA) and National Science, Research and Innovation Fund (NSRF via PMU-B B05F650021), Thailand;
Turkish Energy, Nuclear and Mineral Research Agency (TENMAK), Turkey;
National Academy of  Sciences of Ukraine, Ukraine;
Science and Technology Facilities Council (STFC), United Kingdom;
National Science Foundation of the United States of America (NSF) and United States Department of Energy, Office of Nuclear Physics (DOE NP), United States of America.
In addition, individual groups or members have received support from:
Czech Science Foundation (grant no. 23-07499S), Czech Republic;
FORTE project, reg.\ no.\ CZ.02.01.01/00/22\_008/0004632, Czech Republic, co-funded by the European Union, Czech Republic;
European Research Council (grant no. 101220549), European Union;
Deutsche Forschungs Gemeinschaft (DFG, German Research Foundation) ``Neutrinos and Dark Matter in Astro- and Particle Physics'' (grant no. SFB 1258), Germany;
CONVECS project, CUP C97H23001700002 FESR 2021-2027 program, Italy.

\end{acknowledgement}

%%%%%%%% Bibliography 
\bibliographystyle{utphys}   % Remember we use title in the biblio
\bibliography{bibliography.bib}
%\input {bibliography.tex}  

%%%%%%%%%%%%%%%%%%%%%%%%%%%%%%%%))
% Appendices: yours (if any) + authorlist
%%%%%%%%%%%%%%%%%%%%%%%%%%%%%%%%
\newpage
\appendix

\section{The ALICE Collaboration}
\label{app:collab}
% ALICE Collaboration author list for 2026-10-01
\begin{flushleft} 
\small

D.A.H.~Abdallah\,\orcidlink{0000-0003-4768-2718}\,$^{\rm 133}$, 
M.M.~Abedin$^{\rm 4}$, 
I.J.~Abualrob\,\orcidlink{0009-0005-3519-5631}\,$^{\rm 110}$, 
S.~Acharya\,\orcidlink{0000-0002-9213-5329}\,$^{\rm 49}$, 
A.B.~Adiguzel$^{\rm 90}$, 
K.~Agarwal\,\orcidlink{0000-0001-5781-3393}\,$^{\rm II,}$$^{\rm 23}$, 
G.~Aglieri Rinella\,\orcidlink{0000-0002-9611-3696}\,$^{\rm 32}$, 
L.~Aglietta\,\orcidlink{0009-0003-0763-6802}\,$^{\rm 24}$, 
N.~Agrawal\,\orcidlink{0000-0003-0348-9836}\,$^{\rm 25}$, 
Z.~Ahammed\,\orcidlink{0000-0001-5241-7412}\,$^{\rm 131}$
% ...

\section*{Affiliation Notes}

$^{\rm I}$ Deceased\\
$^{\rm II}$ Also at: INFN Trieste, Trieste, Italy\\
$^{\rm III}$ Also at: Fondazione Bruno Kessler (FBK), Trento, Italy\\
$^{\rm IV}$ Also at: Czech Technical University in Prague, Prague, Czech Republic\\
$^{\rm V}$ Also at: Dipartimento DET del Politecnico di Torino, Turin, Italy\\
$^{\rm VI}$ Also at: University College of Dublin, Dublin, Ireland\\
$^{\rm VII}$ Also at: Department of Applied Physics, Aligarh Muslim University, Aligarh, India\\
$^{\rm VIII}$ Also at: Institute of Theoretical Physics, University of Wroclaw, Wroclaw, Poland\\
$^{\rm IX}$ Also at: Facultad de Ciencias, Universidad Nacional Aut\'{o}noma de M\'{e}xico, Mexico City, Mexico\\
$^{\rm X}$ Also at: Lovely Professional University, Punjab, India
\\

\section*{Collaboration Institutes}

$^{1}$ A.I. Alikhanyan National Science Laboratory (Yerevan Physics Institute) Foundation, Yerevan, Armenia\\
$^{2}$ AGH University of Krakow, Cracow, Poland\\
$^{3}$ Bogolyubov Institute for Theoretical Physics, National Academy of Sciences of Ukraine, Kyiv, Ukraine\\
$^{4}$ Bose Institute, Department of Physics  and Centre for Astroparticle Physics and Space Science (CAPSS), Kolkata, India\\
$^{5}$ California Polytechnic State University, San Luis Obispo, California, United States\\
$^{6}$ Central China Normal University, Wuhan, China\\
$^{7}$ Centro de Aplicaciones Tecnol\'{o}gicas y Desarrollo Nuclear (CEADEN), Havana, Cuba\\
$^{8}$ Centro de Investigaci\'{o}n y de Estudios Avanzados (CINVESTAV), Mexico City and M\'{e}rida, Mexico\\
$^{9}$ Chicago State University, Chicago, Illinois, United States\\
$^{10}$ China Nuclear Data Center, China Institute of Atomic Energy, Beijing, China\\
$^{11}$ China University of Geosciences, Wuhan, China\\
$^{12}$ Chungbuk National University, Cheongju, Republic of Korea\\
$^{13}$ Comenius University Bratislava, Faculty of Mathematics, Physics and Informatics, Bratislava, Slovak Republic\\
$^{14}$ Creighton University, Omaha, Nebraska, United States\\
$^{15}$ Department of Physics, Aligarh Muslim University, Aligarh, India\\
$^{16}$ Department of Physics, Pusan National University, Pusan, Republic of Korea\\
$^{17}$ Department of Physics, Sejong University, Seoul, Republic of Korea\\
$^{18}$ Department of Physics, University of California, Berkeley, California, United States\\
$^{19}$ Department of Physics, University of Oslo, Oslo, Norway\\
$^{20}$ Department of Physics and Technology, University of Bergen, Bergen, Norway\\
$^{21}$ Dipartimento di Fisica, Universit\`{a} di Pavia, Pavia, Italy\\
$^{22}$ Dipartimento di Fisica dell'Universit\`{a} and Sezione INFN, Cagliari, Italy\\
$^{23}$ Dipartimento di Fisica dell'Universit\`{a} and Sezione INFN, Trieste, Italy\\
$^{24}$ Dipartimento di Fisica dell'Universit\`{a} and Sezione INFN, Turin, Italy\\
$^{25}$ Dipartimento di Fisica e Astronomia dell'Universit\`{a} and Sezione INFN, Bologna, Italy\\
$^{26}$ Dipartimento di Fisica e Astronomia dell'Universit\`{a} and Sezione INFN, Catania, Italy\\
$^{27}$ Dipartimento di Fisica e Astronomia dell'Universit\`{a} and Sezione INFN, Padova, Italy\\
$^{28}$ Dipartimento di Fisica `E.R.~Caianiello' dell'Universit\`{a} and Gruppo Collegato INFN, Salerno, Italy\\
$^{29}$ Dipartimento DISAT del Politecnico and Sezione INFN, Turin, Italy\\
$^{30}$ Dipartimento di Scienze MIFT, Universit\`{a} di Messina, Messina, Italy\\
$^{31}$ Dipartimento Interateneo di Fisica `M.~Merlin' and Sezione INFN, Bari, Italy\\
$^{32}$ European Organization for Nuclear Research (CERN), Geneva, Switzerland\\
$^{33}$ Faculty of Electrical Engineering, Mechanical Engineering and Naval Architecture, University of Split, Split, Croatia\\
$^{34}$ Faculty of Nuclear Sciences and Physical Engineering, Czech Technical University in Prague, Prague, Czech Republic\\
$^{35}$ Faculty of Physics, Sofia University, Sofia, Bulgaria\\
$^{36}$ Faculty of Science, P.J.~\v{S}af\'{a}rik University, Ko\v{s}ice, Slovak Republic\\
$^{37}$ Faculty of Technology, Environmental and Social Sciences, Bergen, Norway\\
$^{38}$ Frankfurt Institute for Advanced Studies, Johann Wolfgang Goethe-Universit\"{a}t Frankfurt, Frankfurt, Germany\\
$^{39}$ Fudan University, Shanghai, China\\
$^{40}$ Gauhati University, Department of Physics, Guwahati, India\\
$^{41}$ Helmholtz-Institut f\"{u}r Strahlen- und Kernphysik, Rheinische Friedrich-Wilhelms-Universit\"{a}t Bonn, Bonn, Germany\\
$^{42}$ Helsinki Institute of Physics (HIP), Helsinki, Finland\\
$^{43}$ High Energy Physics Group,  Universidad Aut\'{o}noma de Puebla, Puebla, Mexico\\
$^{44}$ Horia Hulubei National Institute of Physics and Nuclear Engineering, Bucharest, Romania\\
$^{45}$ HUN-REN Wigner Research Centre for Physics, Budapest, Hungary\\
$^{46}$ Indian Institute of Technology Bombay (IIT), Mumbai, India\\
$^{47}$ Indian Institute of Technology Indore, Indore, India\\
$^{48}$ INFN, Laboratori Nazionali di Frascati, Frascati, Italy\\
$^{49}$ INFN, Sezione di Bari, Bari, Italy\\
$^{50}$ INFN, Sezione di Bologna, Bologna, Italy\\
$^{51}$ INFN, Sezione di Cagliari, Cagliari, Italy\\
$^{52}$ INFN, Sezione di Catania, Catania, Italy\\
$^{53}$ INFN, Sezione di Padova, Padova, Italy\\
$^{54}$ INFN, Sezione di Pavia, Pavia, Italy\\
$^{55}$ INFN, Sezione di Torino, Turin, Italy\\
$^{56}$ INFN, Sezione di Trieste, Trieste, Italy\\
$^{57}$ Inha University, Incheon, Republic of Korea\\
$^{58}$ Institute for Gravitational and Subatomic Physics (GRASP), Utrecht University/Nikhef, Utrecht, Netherlands\\
$^{59}$ Institute of Experimental Physics, Slovak Academy of Sciences, Ko\v{s}ice, Slovak Republic\\
$^{60}$ Institute of Physics of the Czech Academy of Sciences, Prague, Czech Republic\\
$^{61}$ Institute of Space Science (ISS), Bucharest, Romania\\
$^{62}$ Institut f\"{u}r Kernphysik, Johann Wolfgang Goethe-Universit\"{a}t Frankfurt, Frankfurt, Germany\\
$^{63}$ Instituto de Ciencias Nucleares, Universidad Nacional Aut\'{o}noma de M\'{e}xico, Mexico City, Mexico\\
$^{64}$ Instituto de F\'{i}sica, Universidade Federal do Rio Grande do Sul (UFRGS), Porto Alegre, Brazil\\
$^{65}$ Instituto de F\'{\i}sica, Universidad Nacional Aut\'{o}noma de M\'{e}xico, Mexico City, Mexico\\
$^{66}$ iThemba LABS, National Research Foundation, Somerset West, South Africa\\
$^{67}$ Jeonbuk National University, Jeonju, Republic of Korea\\
$^{68}$ Korea Institute of Science and Technology Information, Daejeon, Republic of Korea\\
$^{69}$ Laboratoire de Physique Subatomique et de Cosmologie, Universit\'{e} Grenoble-Alpes, CNRS-IN2P3, Grenoble, France\\
$^{70}$ Lawrence Berkeley National Laboratory, Berkeley, California, United States\\
$^{71}$ Lund University Department of Physics, Division of Particle Physics, Lund, Sweden\\
$^{72}$ Marietta Blau Institute, Vienna, Austria\\
$^{73}$ Nagasaki Institute of Applied Science, Nagasaki, Japan\\
$^{74}$ Nara Women{'}s University (NWU), Nara, Japan\\
$^{75}$ National Centre for Nuclear Research, Warsaw, Poland\\
$^{76}$ National Institute of Science Education and Research, Homi Bhabha National Institute, Jatni, India\\
$^{77}$ National Nuclear Research Center, Baku, Azerbaijan\\
$^{78}$ National Research and Innovation Agency - BRIN, Jakarta, Indonesia\\
$^{79}$ Niels Bohr Institute, University of Copenhagen, Copenhagen, Denmark\\
$^{80}$ Nikhef, National institute for subatomic physics, Amsterdam, Netherlands\\
$^{81}$ Nuclear Physics Group, STFC Daresbury Laboratory, Daresbury, United Kingdom\\
$^{82}$ Nuclear Physics Institute of the Czech Academy of Sciences, Husinec-\v{R}e\v{z}, Czech Republic\\
$^{83}$ Oak Ridge National Laboratory, Oak Ridge, Tennessee, United States\\
$^{84}$ Ohio State University, Columbus, Ohio, United States\\
$^{85}$ Physics Department, Panjab University, Chandigarh, India\\
$^{86}$ Physics Department, University of Jammu, Jammu, India\\
$^{87}$ Physics Program and International Institute for Sustainability with Knotted Chiral Meta Matter (WPI-SKCM$^{2}$), Hiroshima University, Hiroshima, Japan\\
$^{88}$ Physikalisches Institut, Eberhard-Karls-Universit\"{a}t T\"{u}bingen, T\"{u}bingen, Germany\\
$^{89}$ Physikalisches Institut, Ruprecht-Karls-Universit\"{a}t Heidelberg, Heidelberg, Germany\\
$^{90}$ Physik Department, Technische Universit\"{a}t M\"{u}nchen, Munich, Germany\\
$^{91}$ Politecnico di Bari and Sezione INFN, Bari, Italy\\
$^{92}$ Research Division and ExtreMe Matter Institute EMMI, GSI Helmholtzzentrum f\"ur Schwerionenforschung GmbH, Darmstadt, Germany\\
$^{93}$ Saga University, Saga, Japan\\
$^{94}$ Saha Institute of Nuclear Physics, Homi Bhabha National Institute, Kolkata, India\\
$^{95}$ School of Physics and Astronomy, University of Birmingham, Birmingham, United Kingdom\\
$^{96}$ Secci\'{o}n F\'{\i}sica, Departamento de Ciencias, Pontificia Universidad Cat\'{o}lica del Per\'{u}, Lima, Peru\\
$^{97}$ SUBATECH, IMT Atlantique, Nantes Universit\'{e}, CNRS-IN2P3, Nantes, France\\
$^{98}$ Sungkyunkwan University, Suwon City, Republic of Korea\\
$^{99}$ Suranaree University of Technology, Nakhon Ratchasima, Thailand\\
$^{100}$ Technical University of Ko\v{s}ice, Ko\v{s}ice, Slovak Republic\\
$^{101}$ The Henryk Niewodniczanski Institute of Nuclear Physics, Polish Academy of Sciences, Cracow, Poland\\
$^{102}$ The University of Texas at Austin, Austin, Texas, United States\\
$^{103}$ Universidad Aut\'{o}noma de Sinaloa, Culiac\'{a}n, Mexico\\
$^{104}$ Universidade de S\~{a}o Paulo (USP), S\~{a}o Paulo, Brazil\\
$^{105}$ Universidade Estadual de Campinas (UNICAMP), Campinas, Brazil\\
$^{106}$ Universidade Federal do ABC, Santo Andre, Brazil\\
$^{107}$ Universitatea Nationala de Stiinta si Tehnologie Politehnica Bucuresti, Bucharest, Romania\\
$^{108}$ University of Cape Town, Cape Town, South Africa\\
$^{109}$ University of Derby, Derby, United Kingdom\\
$^{110}$ University of Houston, Houston, Texas, United States\\
$^{111}$ University of Jyv\"{a}skyl\"{a}, Jyv\"{a}skyl\"{a}, Finland\\
$^{112}$ University of Kansas, Lawrence, Kansas, United States\\
$^{113}$ University of Liverpool, Liverpool, United Kingdom\\
$^{114}$ University of Science and Technology of China, Hefei, China\\
$^{115}$ University of Silesia in Katowice, Katowice, Poland\\
$^{116}$ University of South-Eastern Norway, Kongsberg, Norway\\
$^{117}$ University of Tennessee, Knoxville, Tennessee, United States\\
$^{118}$ University of the Witwatersrand, Johannesburg, South Africa\\
$^{119}$ University of Tokyo, Tokyo, Japan\\
$^{120}$ University of Tsukuba, Tsukuba, Japan\\
$^{121}$ University of Zagreb Faculty of Science, Department of Physics, Zagreb, Croatia\\
$^{122}$ Universit\"{a}t M\"{u}nster, Institut f\"{u}r Kernphysik, M\"{u}nster, Germany\\
$^{123}$ Universit\'{e} Clermont Auvergne, CNRS/IN2P3, LPC, Clermont-Ferrand, France\\
$^{124}$ Universit\'{e} de Lyon, CNRS/IN2P3, Institut de Physique des 2 Infinis de Lyon, Lyon, France\\
$^{125}$ Universit\'{e} de Strasbourg, CNRS, IPHC UMR 7178, F-67000 Strasbourg, France, Strasbourg, France\\
$^{126}$ Universit\'{e} Paris-Saclay, Centre d'Etudes de Saclay (CEA), IRFU, D\'{e}partment de Physique Nucl\'{e}aire (DPhN), Saclay, France\\
$^{127}$ Universit\'{e}  Paris-Saclay, CNRS/IN2P3, IJCLab, Orsay, France\\
$^{128}$ Universit\`{a} degli Studi di Foggia, Foggia, Italy\\
$^{129}$ Universit\`{a} del Piemonte Orientale, Vercelli, Italy\\
$^{130}$ Universit\`{a} di Brescia, Brescia, Italy\\
$^{131}$ Variable Energy Cyclotron Centre, Homi Bhabha National Institute, Kolkata, India\\
$^{132}$ Warsaw University of Technology, Warsaw, Poland\\
$^{133}$ Wayne State University, Detroit, Michigan, United States\\
$^{134}$ Yale University, New Haven, Connecticut, United States\\
$^{135}$ Yildiz Technical University, Istanbul, Turkey\\
$^{136}$ Yonsei University, Seoul, Republic of Korea\\
$^{137}$ Affiliated with an institute formerly covered by a cooperation agreement with CERN\\
$^{138}$ Affiliated with an international laboratory covered by a cooperation agreement with CERN.\\

\end{flushleft} 
  
\end{document}